# Element selection for functional materials discovery by integrated machine learning of elemental contributions to properties

A. Vasylenko[1], D. Antypov[1], V. Gusev[1], M. W. Gaultois[1], M. S. Dyer[1], M. J. Rosseinsky[1*]

[1] Department of Chemistry, University of Liverpool, Crown Street L6 97ZD, UK

*corresponding author

## Abstract

Fundamental differences between materials originate from the unique nature of their constituent chemical elements. Before specific differences emerge according to the precise ratios of elements (the composition) in a given crystal structure (a phase), a material can be represented by its phase field defined simply as the set of its constituent chemical elements.

By working at the level of the periodic table, assessment of materials at the level of their phase fields reduces the combinatorial complexity to accelerate screening, and circumvents the challenges associated with composition-level approaches – poor extrapolation within phase fields, and the impossibility of exhaustive sampling. This early stage discrimination combined with evaluation of novelty of phase fields aligns with the outstanding experimental challenge of identifying new areas of chemistry to investigate, by prioritising which elements to combine in a reaction.

Here, we demonstrate that phase fields can be assessed with respect to the maximum expected value of a target functional property and ranked according to chemical novelty.

We develop and present PhaseSelect, an end-to-end machine learning model that combines the representation, classification, regression and ranking of phase fields. First, PhaseSelect constructs elemental characteristics from the co-occurrence of chemical elements in computationally and experimentally reported materials, then it employs attention mechanisms to learn representation for phase fields and assess their functional performance. At the level of the periodic table, PhaseSelect quantifies the probability of observing a functional property, estimates its value within a phase field and also ranks a phase field's novelty, which we demonstrate with significant accuracy for three avenues of materials' applications: high-temperature superconductivity, high-temperature magnetism, and targeted bandgap energy.

## Introduction

Conceptualization of novel materials begins at the level of the periodic table with selection of chemical elements for synthetic investigation. There is a variety of possible ratios or compositions that can be formed from a set of chemical elements (*e.g.*, {Cu, O, Br}) leading to different materials (phases, *e.g.*, $Cu_2BrO_3$); the field of these potential realizations can be defined as a phase field (*e.g.*, {Cu, O, Br}). The choice of phase field to investigate ultimately determines the outcome of the synthetic work and the functional properties of the prospective materials. Hence, this high-level discrimination of which phase field to investigate is essential before significant resources are committed to the investigation of individual compositions within a phase field.

The fundamental differences between chemical elements result in a gamut of material properties in thousands of compositions accumulated in materials databases[1–3]. These data have been exploited by a surge of machine learning (ML) methods aiming to predict material properties from the knowledge of their compositions and structures[4,5]. Both structure- and composition-based approaches demonstrate powerful capabilities of ML for acceleration of materials discovery[6–8]. Searching for truly new materials, for which neither the composition nor the crystal structure is known beforehand, and open-ended approaches of curious formulations[9], generative approaches[10] and serendipity based recommender systems[11] have been applied to navigate experimentation in the uncharted chemistry spaces. Exhaustive enumeration of unexplored compositions is impossible, making existing ML approaches based on extensive screening susceptible to missing a potential new material. Moreover, as the quality of ML models is heavily dependent on the available data for training, composition-based models inherit the historical bias of the past research preferences towards particular materials families, such as extensive studies of cuprates as superconductors[12]. The imbalanced datasets are known to have detrimental effects on the model performance and the capability to extrapolate the patterns of composition-property relationships into unexplored chemistry[12–15].

This highlights the need for the evaluation of materials at the high governing level of the constituent elements, prior to their compositional assessment. By aggregating materials into the phase fields, one retains the fundamental differences between the sets of chemical elements, while eliminating the risk of missing promising compositions. Additionally, by consolidating compositions into phase fields, the presence of different material families in historical data is redistributed: data balance is improved, with materials represented uniformly in the datasets. This improves ML model accuracy and capability to extrapolate composition-property relationships into uncharted chemistry[14]. This phase field level of approach has already shown merit, with the experimental realization of new stable materials in phase fields prioritised using similar methods[16].

In this work, our goal is to assess the attractiveness of unexplored candidate inorganic functional materials at the level of the periodic table by identifying unexplored phase fields that are likely to contain these candidates. This circumvents the combinatorial challenge of exhaustive individual assessment of all possible compositions built from the chosen elements. Further, this workflow assists decision-making in new areas of experimental solid-state inorganic chemistry, by prioritising which elements to combine in a reaction; currently, this is the only ML tool addressing this challenge. The high-level prioritisation aims to provide a computationally undemanding guide for research at its earliest stage, which is only a part of the total materials discovery challenge, as there are undiscovered phases within phase fields that have been explored. The proposed unexplored phase fields prioritisation can be followed by more computation- and data-intensive investigations of materials. The guidance is broadly applicable to unexplored inorganic materials, where change of constituent elements plays a determining role for stability and function; the same approach does not apply to prioritisation of synthetic routes, nor in organic chemistry, which may be more suitably addressed by other computational guides[17,18].

We develop an end-to-end integrated (from elemental representation to the phase fields assessment) machine learning approach (PhaseSelect) that can prioritise phase fields with respect to both functional performance (*e.g.*, maximum value of a target property within a phase field) and chemical similarity (*i.e.*, similarity to phase fields with stable compositions).

PhaseSelect starts with semi-supervised learning of representations for chemical elements from the elemental co-occurrence in all calculated and experimentally reported materials (inspired by the approach in [19]) coupled with the supervised assessment of materials' functional performance – regression and binary classification. The coupling is achieved through the 'attention' weighting of the contributions of constituent chemical elements to the functional performance of the material. The attention mechanism originates from computer science research in natural language representation[20], and is implemented in our model to learn the elemental representations that align best with the resulting representation of a phase field, such that its functional performance is quantified most accurately.

We demonstrate the predictive power of PhaseSelect in quantitative assessments of phase fields with three functional properties of interest: superconducting transition temperature, Curie temperature, and band gap energy. The models for each dataset are trained independently;

composition-level data with associated properties from SuperCon[3] and/or Materials Platform for Data Science (MPDS)[1] databases is first aggregated into collections of phase fields, and each phase field is labelled according to the maximum reported value of all materials within it.

In a regression task, we verify a) viability of the materials description simply as elemental sets and b) PhaseSelect' capability to learn informative phase fields representations while predicting a maximum value achievable within phase fields. In binary classification, we discriminate materials with respect to performance thresholds we define for each property ($T_c$=10K, $T_C$ =300 K, $E_{Gap}$=4.5eV) that reflect practical interests in high-temperature superconductors, magnetic materials, and dielectrics/ruling out candidates for photovoltaics. In both regression and classification, PhaseSelect demonstrates significant improvement of performance in comparison to the baseline model – random forest[21] with Magpie descriptors of elements[22]. We combine binary classification with regression to first, assign phase fields to low- or high-performing classes of materials and then predict maximum expected values of property of interest. This develops reliable quantitative metrics for fast high-level discrimination and screening of materials phase fields at scale.

The phase field representations constructed during property classification are used further for unsupervised learning of similarity between elemental combinations in materials databases that afford stable compositions. This stage completes the end-to-end assessment of elemental combinations by producing the ranking of chemical novelty for unexplored phase fields.

The arising metrics of the phase fields – functional performance (quantified by regression and the classifier probability of belonging to high-performance class) and chemical novelty (quantified by distance in representation space from phase fields with stable compositions) – can be coupled or used independently for any combination of elements, creating a map of potentially attractive phase fields for future research. This can provide quantitative guidance to human researchers in the consequential and costly choice of phase fields for investigation and discovery of novel functional materials.

## Results and discussion

### PhaseSelect model architecture

At the level of the phase fields, relationships between elemental combinations and their synthetic accessibility have been studied with unsupervised machine learning and validated experimentally[16]. Here, we employ an integrated statistical description of chemical elements and their combinations to learn what elemental combinations have high probabilities of both novelty and high values of target properties. PhaseSelect architecture combines several artificial neural networks (ANN) that are trained end-to-end as an integrated model:

$$S: \mathbb{R}^m \to \mathbb{R}^1, \quad S(\mathbf{p}) = R \circ W \circ A(\mathbf{C})\mathbf{p}, \tag{1.1}$$

$$P: \mathbb{R}^m \to \mathbb{R}^1, \quad P(\mathbf{p}) = P(\overline{\overline{\mathbf{p}}}), \tag{1.2}$$

where $S$ is a supervised model for classification (regression), $P$ is unsupervised model for chemical similarity ranking; a phase field, $\mathbf{p}$, of dimensionality $m$ (e.g., $m$ = 3 for ternary) is encoded via semi-supervised model, $A$, which learns representations for chemical elements from the matrix of elemental co-occurrence, $\mathbf{C}$, in all calculated and experimentally reported materials (inspired by the approach in [19]); representation learning is guided by supervised assessment (classification or regression), $R$, of materials' functional performance; learnt representations for phase fields, $\overline{\overline{\mathbf{p}}} = W \circ A(\mathbf{C})\mathbf{p}$, minimise error of $R$ and can be further used for unsupervised learning, $P$, of chemical similarity (and inversely, novelty) of phase fields (1.2); the coupling between elemental and phase field representations is achieved through the 'attention' weighting, $W$, of the contributions of constituent chemical elements to the functional performance of the material; by symbols $\circ$ and parenthesis in Eq. 1.1, 1.2 we denote connectivity of the data flow and transformation of data. The architecture of the model is illustrated in Figure 1.

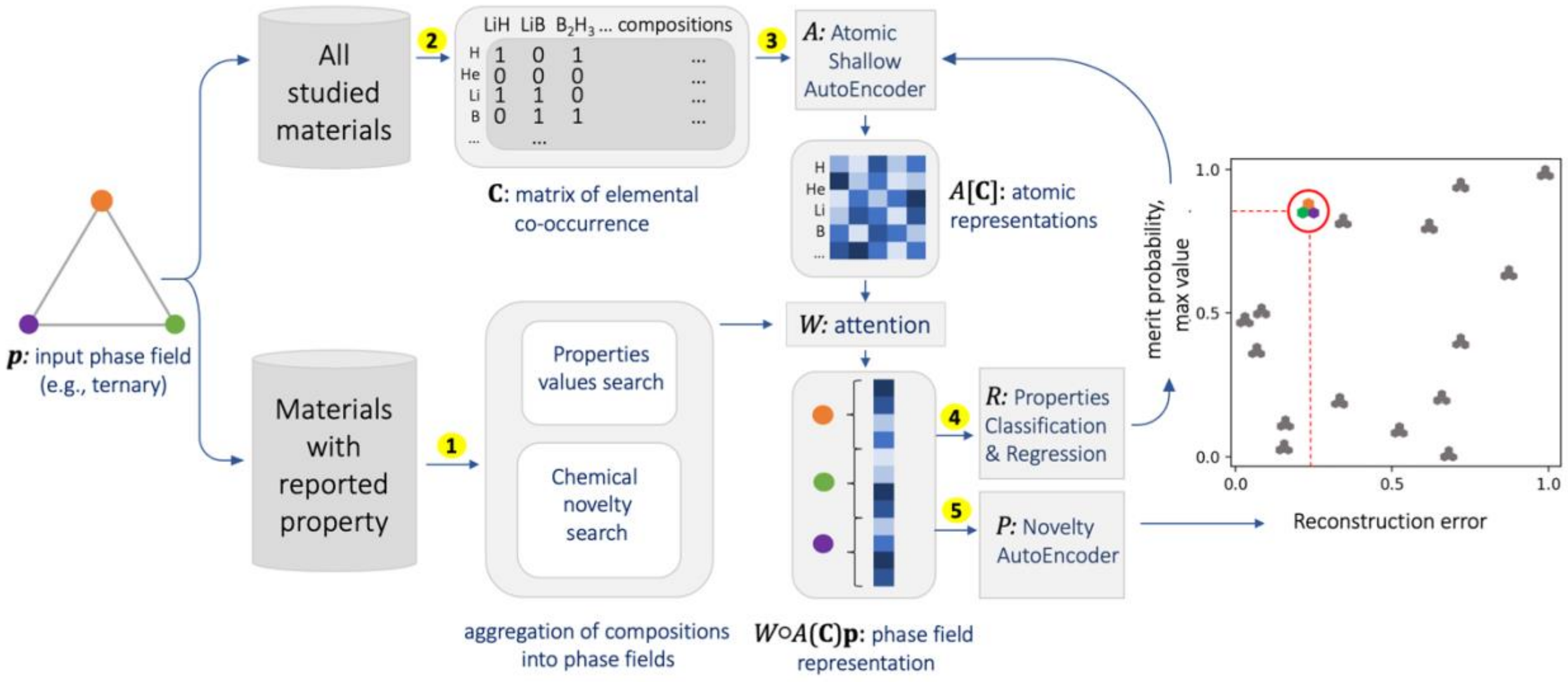

**Figure 1. PhaseSelect predicts properties and chemical accessibility of phase fields. Model architecture.** Arrows show the information flow between the various components described in this paper: 1) experimentally confirmed compositions are aggregated into the phase fields; the maximum values of the properties in the phase fields are selected; 2) compositional environments (elemental co-occurrence in materials) are aggregated from all theoretically and experimentally studied materials; 3) unsupervised learning of elemental representation from data collected in 2); 4) supervised classification of phase fields by maximum achievable values of the properties; the predicted probability of entering the high-value class is used as a probability of high functional performance (merit probability); regression to maximum value achieved within phase fields 5) unsupervised ranking of the phase fields by similarity to synthetically stable materials; metrics derived in 4) and 5) result in a map of the phase fields' likelihood to form stable compounds with desired properties. The model is trained end-to-end so the losses of learning the elemental representation (3) and classification (4) are minimised simultaneously.

PhaseSelect consists of several connected modules (ANN depicted as the sharp-corner rectangles in Figure 1) that pass information from the databases (dark grey cylinders in Figure 1), while transforming the data (different data representations are depicted as the rounded-corner windows in Figure 1) and are trained simultaneously while minimising the compound loss. We describe the data processing and the mechanisms of these modules in the following sections.

## Aggregation of compositions into phase fields

For assessment of materials at the level of phase fields (See bottom stream in Figure 1) we process the materials databases, where experimentally verified values of the target property are reported for a large number of compositions[1,3]. Some historical bias is present in every representation of materials data. By representing materials as phase fields, the bias can be decreased via uniform representation of all phase fields as described below. Materials built from the same constituent elements are aggregated into one phase field, with the associated property value corresponding to the maximum reported property value among all reported materials within this phase field. For example, in the SuperCon database, there are many compositions reported in the Y-Ba-Cu-O phase field with a high critical temperature, including $YBa_2Cu_3O_7$ ($T_c$ = 93 K) and $Y_3Ba_5Cu_8O_{18}$ ($T_c$ = 100.1 K) – the highest reported temperature in Y-Ba-Cu-O. Hence, Y-Ba-Cu-O enters the data for training our classification model for superconductors with 100.1 K as the corresponding maximum value. Aggregation of materials with reported superconducting transition temperature, Curie temperature and energy bandgap forms three datasets with 4826, 4753 and 40452 phase fields respectively. Division of the datasets into two classes by the threshold values for the corresponding properties – 10 K, 300 K and 4.5 eV for superconducting transition temperature, Curie temperature and energy bandgap, respectively – forms reasonably balanced data classes with 3311:1515, 2726:2027 and 20910:19690 phase fields, respectively, with data distributions illustrated in Figure 2a-c. The data balance is further improved by class-weighting in the corresponding classification models[23]. The rapidly decreasing number of explored phase fields with reported superconducting properties at temperatures above 10 K (See Figure 2b) proves development of reliable models for classification with respect to temperatures higher than 10 K particularly challenging (See Supplementary Fig. 1)[24]. Nevertheless, despite the broad aggregation of high-temperature superconducting materials into a single class (with $T_c$ > 10 K), accurate classification of unexplored materials into the two classes divided by the chosen threshold value would allow fast screening for novel high-temperature superconductors. Similarly, a binary classification enables a fast screening of novel materials for applications as high-temperature magnetic materials and targeted bandgap materials.

Across the three property datasets, the phase fields are formed from up to 12 constituent elements, with the majority of data represented by ternary, quaternary and quinary phase fields (See Figure 2d). The abundance of chemical elements among the explored materials in the databases is illustrated in Figure 2e. All datasets have similar trends with peaks for materials containing, e.g., carbon, oxygen, sulphur, with an especially pronounced match between elemental distribution in datasets with materials for superconducting and magnetic applications (See inset in Figure 2e). The data distributions across different chemical elements observed in Figure 2e reflect the biases in the input data: e.g., magnetism is associated with Fe predominantly, while superconductivity with Cu, etc.

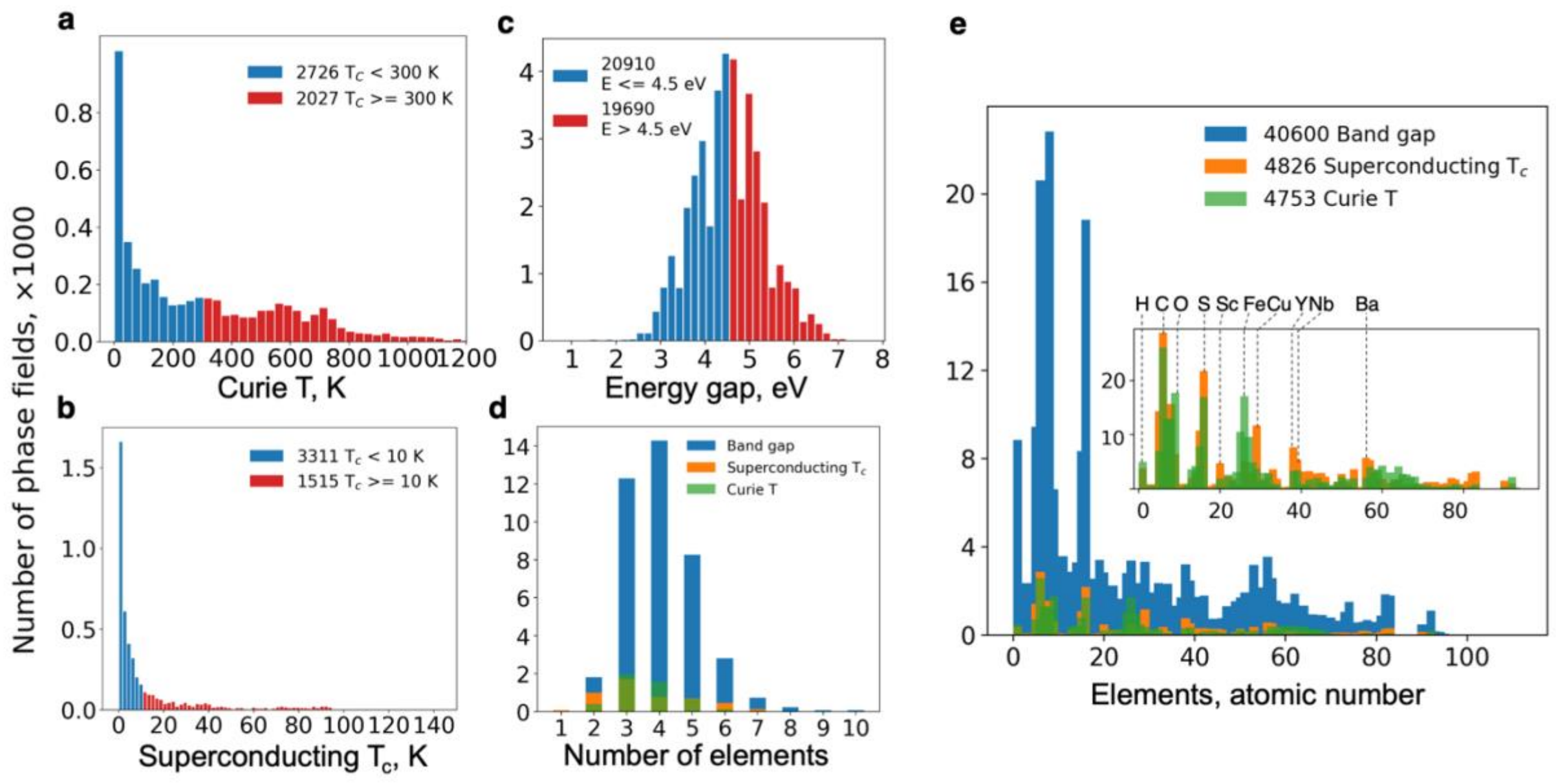

**Figure 2. Aggregation of compositions into phase fields. a** Distribution of phase fields of magnetic materials in MPDS[1] with respect to the maximum associated Curie temperature $T_C$. The materials' classes "low-

temperature" and "high-temperature" magnets are divided at $T_C$ = 300 K as 2726:2027 phase fields. **b** Distribution of phase fields of superconducting materials (joined datasets from SuperCon[3] and MPDS) with respect to the maximum associated superconducting transition temperature $T_c$. The materials' classes "low-temperature" and "high-temperature" superconductors are divided around $T_c$ = 10 K as 3311:1515 phase fields. **c** Distribution of phase fields of materials with reported value of energy gap in MPDS with respect to the maximum associated band gap. The materials' classes "small-gap" and "large-gap" are divided around E = 4.5 eV as 20910:19690 phase fields. **d** Distributions of materials with respect to the number of constituent elements are similar for all datasets: the majority of the reported compositions belong to ternary, quaternary and quinary phase fields. **e** Content of individual chemical elements among the explored materials in the databases; the total numbers of phase fields in the corresponding datasets are given in the legend. All datasets have similar trends with pronounced peaks for materials containing, e.g. carbon, oxygen, silicon. The inset illustrates overlap in trends for elemental distribution in explored materials for superconducting and magnetic applications, where the peaks of the prevalent constituent elements are highlighted.

Description of materials as sets of constituent chemical elements should help mitigate the biases in the data accumulated over time due to the focused studies of particular families of materials: both understudied (e.g., single composition) and established phase fields (e.g., multiple compositions, solid solutions, etc. ) are uniformly represented as sets of elements; however phase field representation of materials cannot completely eliminate the bias related to factors including historical interest, availability of particular chemical elements and similarity of the chemistry studied to that of minerals (Figure 2e). A historical trend in materials research may be also suggested by weak correlations of property variance with maximum values and with the number of compositions reported in a phase field (See Supplementary Fig. 16). At this level of description, we ignore the property variance and focus on the maximum values achievable by materials built from selected chemical elements.

## Elemental representation and phase field representation

To learn elemental characteristics from the compositional environments – explored chemical compositions, where the chemical elements are found to form the variety of stable and metastable materials – we build a module for elemental representation based on a large materials database that includes both experimental and theoretical materials[19,25]. For each chemical element one can build a binary vector indicating its presence in chemical formulae in the database. The database is expanded into a table similarly to the approach proposed in reference[19] (depicted as a matrix of coexisting elements and compositional environments in the materials in Figure 1, 2)). The rows of the table correspond to the chemical elements, the columns are the remainders of the compositional formulae of the reported compounds, which we define here as compositional environments. For example, from stability of $Li_3PO_4$ we can learn about its constituent elements, Li, P, O and their compositional environments, "()3PO4", "()Li3O4" and "()4Li3P" respectively. In this notation, empty parentheses denote an element that by combining with the compositional environment forms a composition. Similarly, all alkali metals form the tri-"element" phosphates with "()3PO4", while trivalent elements do not, as they form the one-"element" phosphates with "()PO4" instead. In the proposed matrix representation[19], the intersections of the rows for elements with the columns for compositional environments are filled with ones if the resulting composition is reported in [25] and with zeros otherwise. The resulting sparse matrix, $\mathbf{C}$, represents coexistence of the $n$ chemical elements and $d$ compositional environments in the materials. We then employ a shallow autoencoder neural network – an unsupervised ML technique– to reduce the dimensionality of the matrix $\mathbf{C}$:

$$A: \mathbb{R}^d \to \mathbb{R}^k, \qquad A(\mathbf{C}) = \sigma_{ReLU}\left(\sum_i \omega_i c_i + b_i\right), \tag{2.1}$$

which forms an encoder part of the autoencoder, where $\omega_i$ and $b_i$ are weights and biases ANN, $c_i$ are rows of the matrix $\mathbf{C}$ and $\sigma_{ReLU}$ is ReLU activation function[26]. Autoencoder is employed to condense the information into the rich latent space of dimensionality $k$, in which similar elemental vectors (of length $k$) are grouped close to each other; during the autoencoder training, the encoder $A$ is tuned in conjunction with the mirrored-size decoder, $D$, to minimise the loss $\mathcal{L}(D, A)$:

$$D: \mathbb{R}^k \rightarrow \mathbb{R}^d, \qquad \mathcal{L}(D, A) = \|\mathbf{C} - D \circ A(\mathbf{C})\|^2 , \tag{2.2}$$

which is Euclidian distance between original and reconstructed elemental vectors in matrix $\mathbf{C}$.

We study the effects of the size of dimensionality $k$ of thus derived elemental vectors on the mean absolute error of regression network and classification accuracy to select the most efficient elemental description (Supplementary Fig. 1). We use the vectors of the most efficient latent space as elemental representations to build up the phase fields descriptions as matrices $\widetilde{\mathbf{p}} = A(\mathbf{C})\mathbf{p}$ of size $(m, k)$ , where $m$ is a number of constituent elements in a phase field and rows are the corresponding elemental vectors (Figure 3a).

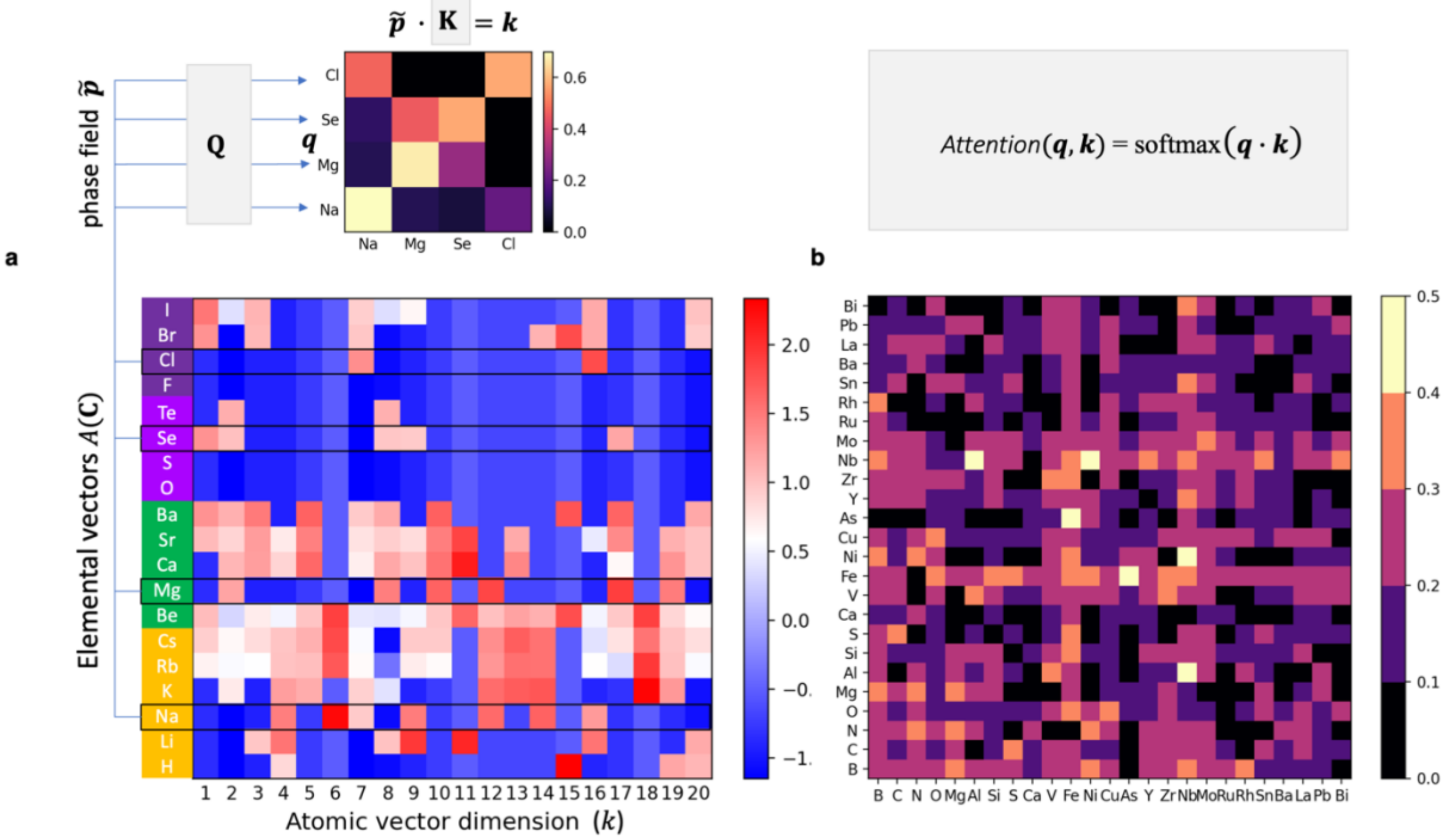

**Figure 3. Elemental representations and their contributions to the phase fields' properties. a** Elemental representation vectors learnt via autoencoder A(**C**) (Eq. 2.1) in $k$ = 20 dimensions for the 1$^{st}$, 2$^{nd}$, 16$^{th}$ and 17$^{th}$ atomic groups of the periodic table. The values (corresponding colour) illustrate differences and correlations between constructed elemental features (vectors' components) in the neighbouring chemical elements and groups. The full stack of elemental vectors for the whole periodic table is extracted by PhaseSelect's elemental

autoencoder shallow neural network, from the sparse matrix of chemical elements and compositional environments built for the Materials Project database[19,25]; for an example unexplored quaternary phase field, Na-Mg-Se-Cl,, the corresponding contributions of the chemical elements to the likelihood of high-temperature superconductivity of this combination are calculated as the attention scores[20] (Supplementary Fig. 2-6). **b** Attention scores – a scaled matrix multiplication of the aspect representation matrices **Q** and **K** of a phase field $\tilde{\boldsymbol{p}}$ – are trained during the fitting of the model for phase fields classification by the target property (Eq. 3). Here, attention to the elemental contributions to superconducting behaviour is visualised: elemental combinations that include e.g., Fe, Nb, Cu, Ni, Mo receive high attention in prediction of high-temperature superconductivity.

To emphasise the differences in the contributions of individual chemical elements to each of the studied phase field's properties, we employ the multi-head local attention[20]:

$$W: \mathbb{R}^m \to \mathbb{R}^m, \;\; W_h(\tilde{\mathbf{p}}) = \sum_i \alpha^h_{\boldsymbol{q}\boldsymbol{k}_i} \boldsymbol{v}^h_{\boldsymbol{k}_i}, \tag{3}$$

where for each head, $h$, attention scores $\alpha^h_{\boldsymbol{q}\boldsymbol{k}_i} = \frac{\exp\left(\boldsymbol{q}^h \cdot \boldsymbol{k}^h_i\right)}{\sum_j \exp\left(\boldsymbol{q}^h_j \cdot \boldsymbol{k}^h_i\right)}$, and $\boldsymbol{q}^h, \; \boldsymbol{k}^h, \; \boldsymbol{v}^h$ are different aspects representations of the phase fields: $\mathbf{Q}^h\tilde{\mathbf{p}}$, $\mathbf{K}^h\tilde{\mathbf{p}}$, $\mathbf{V}^h\tilde{\mathbf{p}}$, respectively. The aspects representations matrices $\mathbf{Q}^h$, $\mathbf{K}^h$, $\mathbf{V}^h$ are trainable parameters that are initialised randomly. The resulting representation is obtained by concatenating the heads representations $W_h$. The attention scores $\alpha^h_{\boldsymbol{q}\boldsymbol{k}_i}$ are the weights for the constituent elemental vectors contributing to the accurate prediction of the phase field values of a targeted property, these weights highlight the intermediate and interpretable results of the ML reasoning process well-aligned with the human understanding of chemistry of materials (Figure 3b, Supplementary Fig. 2-6).

From the calculated attention scores, one can infer elemental, pair and complex many-element contributions to the targeted functional properties. For example, one can obtain a distribution of attention scores for the constituent elements that affect a particular property the most across the phase fields (Supplementary Fig. 6) and the maps for pairs of elements, which contribute the most in co-occurrence (Fig. 3b, Supplementary Fig. 3-5). From the latter, while some pairs may be familiar to human researchers, e.g., Nb-Al, Nb-Ni, Cu-O and Fe-As exhibiting high-temperature superconductivity, other less obvious pairs, e.g., N-Na, La-Cl and P-Si motivate further research of chemistry of superconductivity. Furthermore, more complex correlations in ternary and quaternary phase fields, readily evaluated with our approach, are difficult to visualise and assess by simple statistics, which underlines the additional utility of this approach.

When building a phase field representation for the downstream tasks of property values assessments and chemical similarity ranking, the element attention-weighted phase field matrix $W(\tilde{\mathbf{p}})$ (Eq. 3) is flattened to form a ($m \times k$)-dimensional vector $\bar{\bar{\boldsymbol{p}}}$, where $m$ is a number of constituent elements in a phase field, $k$ is the chosen length of the elemental vector.

## Regression and classification by properties' values and ranking by chemical novelty

Supervised assessment of properties in PhaseSelect is performed by two separate ANN that are a) regressor that predicts targeted property values of phase fields b) classifier that assigns the phase fields to the corresponding classes of the properties' values:

$$R: \mathbb{R}^{m \times k} \to \mathbb{R}^1, \qquad R(\bar{\bar{\mathbf{p}}}) = \sigma_q\left(\sigma_{ReLU} \sum_i [\delta_i \omega_i p_i + b_i]\right), \tag{4}$$

$$\text{a) } \sigma_q(z) = \sum_i \omega_i z_i + b_i, \quad \mathcal{L}(R,t) = MAE(R,t), \tag{4.1}$$

$$\text{b) } \sigma_q(z) = \frac{1}{1+\exp(-z)}, \qquad \mathcal{L}(R,t) = BCE(R,t), \tag{4.2}$$

where $\delta$ stands for a dropout function[27], $\omega_i$ and $b_i$ are weights and biases of ANN nodes, and $p_i$ are elements of the phase field vector; loss functions of prediction $R$ of the target value $t$, $\mathcal{L}(R,t)$, are mean absolute error (MAE) and binary cross entropy (BCE), for regression and classification respectively. The corresponding loss functions are summed up with the representation learning loss $\mathcal{L}(A,D)$ (Eq. 2.2) for simultaneous end-to-end training of representations and predictions: $\mathcal{L}_{compound} = \mathcal{L}(A,D) + \mathcal{L}(R,t)$.

For each dataset, we train an individual regressor with the architecture described above in Equations (1)-(4.1) and Figure 1, in which elemental representations, phase field representations with attention to elemental contributions and predictions of a maximum achievable value for a particular property are trained end-to-end. The relatively small size of the datasets limits the possibility of a prior train-test split, in which 20% of data is withheld from cross-validation without reducing the models' performance due to the lack of training data. Hence, in line with the common benchmarking practice for ANN models for materials science[28], we perform 5-fold cross-validation, while excluding possibility for data leakage by splitting data into non-overlapping folds prior to any data transformation, and report the average MAEs across the folds for dataset (See Supplementary Fig. 7, Table 1). In line with the best practices[29], we compare the performance of PhaseSelect with the baseline models. For the latter we employ random forest[21] regression and classification models, whereas phase fields are described as ($m \times k$)-dimensional vectors, with $k$ Magpie elemental features[22] (See Supplementary Fig. 10). In Figure 4, we illustrate the match between PhaseSelect predictions with true values and an improved performance of PhaseSelect in comparison to the baseline models for all datasets studied.

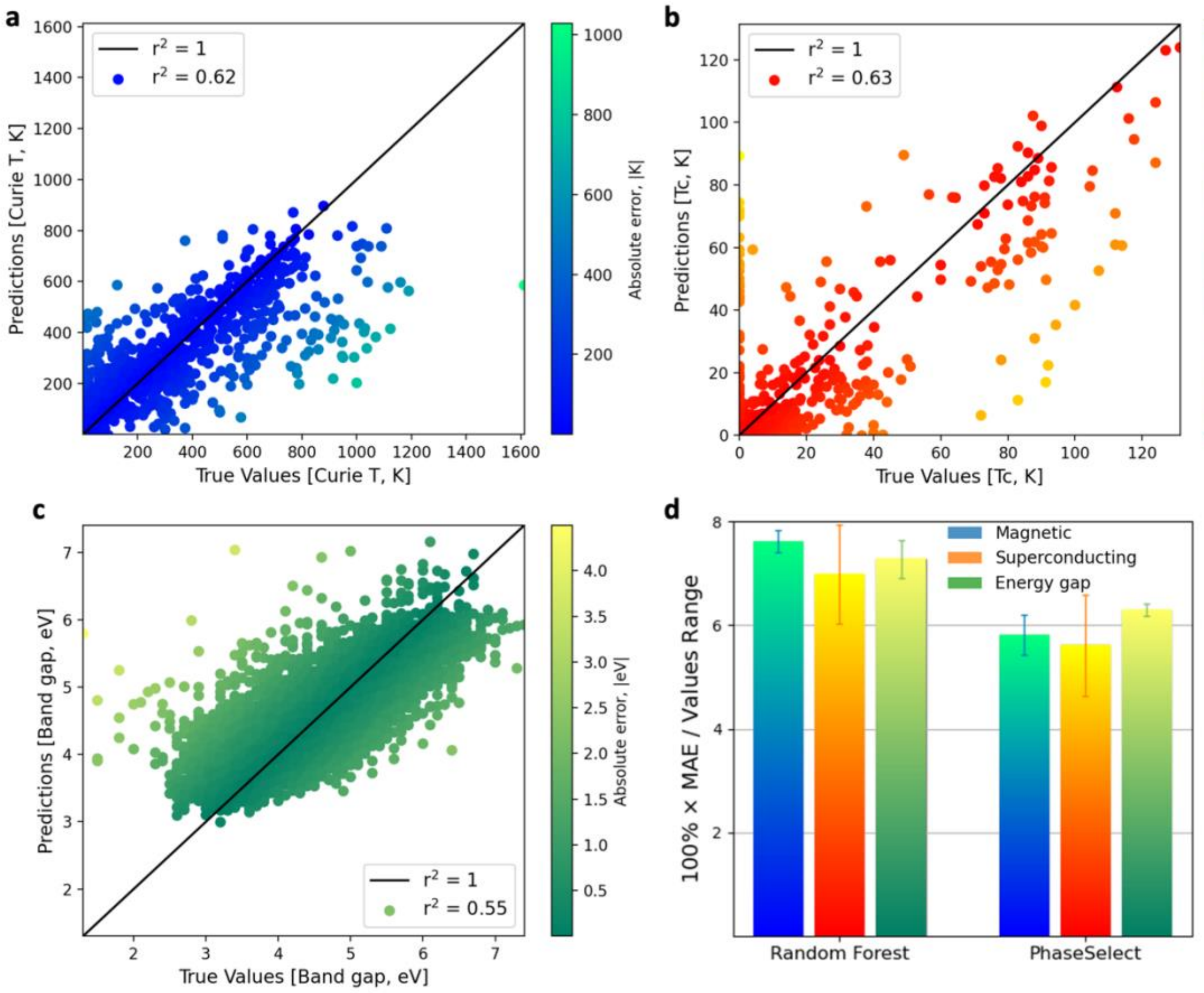

**Figure 4. PhaseSelect regressions of phase fields to targeted properties.** The models are trained and tested on a random 80-20% train-test split of the datasets: MAEs and $r^2$ scores are consistent with the average results in 5-fold cross-validations. **a** Predictions vs true values for maximum Curie temperatures reported and peer-reviewed within phase fields in MPDS[1]; **b** predictions vs true values for maximum superconducting transition temperatures experimentally reported within phase fields in Supercon[3]; **c** predictions vs true values for maximum energy band gap reported and peer-reviewed within phase fields in MPDS[1]; **d** averaged over 5-folds' MAEs scaled with a range of values in the corresponding database for Random Forest[21] with Magpie[22] descriptors (See Supplementary Fig. 10) and PhaseSelect models; the error bars are the standard deviations of MAE achieved in 5 non-overlapping data folds in cross-validation.

The illustrated in Figure 4a-c distributions of values, the corresponding MAE and $r^2$ scores are calculated for the random 20% of data for each property dataset, while PhaseSelect is trained on the remaining 80%; the metrics are characteristic for the results observed in k-fold cross-validations: MAE are within the standard deviations, including scaled MAE highlighted in Figure 4d. Description of phase fields as concatenation of elemental vectors and regression to the maximum values reported within the phase fields is a demonstrated a viable approach by Random Forest (See comparison to dummy model regressions in Supplementary Fig. 9, 10). PhaseSelect approach further improves regression metrics for all studied properties (Figure 4d), while capturing major trends in elemental combinations to property relationships.

For guidance of synthetic combinatorial chemistry and acceleration of screening at scale, it is practically useful to first accurately assign candidates to the major clusters of performance before prioritisation of elements within high-performing group. To ensure high accuracy of assignment we employ binary classification of phase fields to the 'poor' and 'high' performing groups: the phase fields in each dataset are divided into two classes (Figure 2a-c) that are labelled with '1' for the phase fields with associated property values above the chosen thresholds, and with '0' for the remaining phase fields. Three independent classifiers, one for each dataset – for superconducting materials and magnetic materials, and materials with a reported value of energy gap – are trained end-to-end with the architecture described in Equations (1)-(4.2) and Figure 1. Because the elemental characteristic and their relation to the materials properties are learnt from the reported chemistry, where the reports of the negatives (materials not possessing certain properties) are absent, the classification models are not trained to predict manifestation of target properties or their absence. Instead, for the phase fields that may contain compositions with target properties, the classification models predict the probability of reaching high values of these properties within the phase fields. For example, in the training set for the materials with reported values of energy gap, none were reported with zero value (Fig. 2c). To verify the predictive power of the model trained on such data for the energy band gap classification, we have tested all 9816 intermetallic ternaries that do not have energy band gap values reported in MPDS (Supplementary discussion). 99.96% of the intermetallic ternary phase fields were classified as low energy gap materials (<4.5 eV) demonstrating the model's ability to extrapolate chemical patterns of elemental combinations – properties relationships, in absence of the zero-gap examples.

Similarly to the regressors, we study the performance of classification models in the 5-fold cross-validations (Supplementary Fig. 7, Table 1) and report the averaged results over the folds in accord to the common benchmarking practice[28]. The average accuracy across the validation sets is 80.4, 86.2, 75.6 % for classification with respect to superconducting transition temperature, Curie temperature, and energy gap respectively. For the predictive models, we adopt all available data in the three datasets for training. Noting the stochastic nature of the machine learning ANN, we employ averaging of the predicted probabilities over the ensemble of 300 models, this minimises the differences in training processes and derived models' parameters (Supplementary Fig. 13). The ensemble with the minimised variance in predictions enables assessment of the materials' properties not only by the assigned binary classes, that are threshold-dependent (Figure 4d, Supplementary Fig. 12, Supplementary Table 1), but also by the continuous values of probabilities as a measure of likelihood of achieving a desired property value. The latter helps to prioritise the materials for synthesis and further investigation.

A deep AutoEncoder neural network in Equation (1.2) learns patterns of chemical similarity with the experimentally verified materials data. Similar to our original approach reported in [16], an unsupervised de-noising AutoEncoder learns the patterns of similarity in data while reducing dimensionality of the phase fields representations. The training consists of two parts, in principle equivalent to Autoencoder in Equations (2.1)-(2.2): encoding into a reduced dimensionality latent space, where phase fields representations are reorganised, so the similar phase fields are aligned, and decoding from the latent representation into the reconstructed images of original vectors. This reorganisation via the AutoEncoder enables ranking of the phase fields by their reconstruction errors, that reflect differences of individual entries from general patterns in data. Hence, elemental combinations that are unlikely to manifest conventional bonding chemistry, (i.e., combinations nonconforming with synthetically accessible compositions in training data) exhibit high reconstruction errors[16]. By combining the ranking AutoEncoder with the classification of properties, we can transfer some advantages of the supervised learning to unsupervised assessment of similarity between the phase fields. This is achieved by encoding the phase fields with the elemental representations that are learnt in the supervised setting of the end-to-end training.

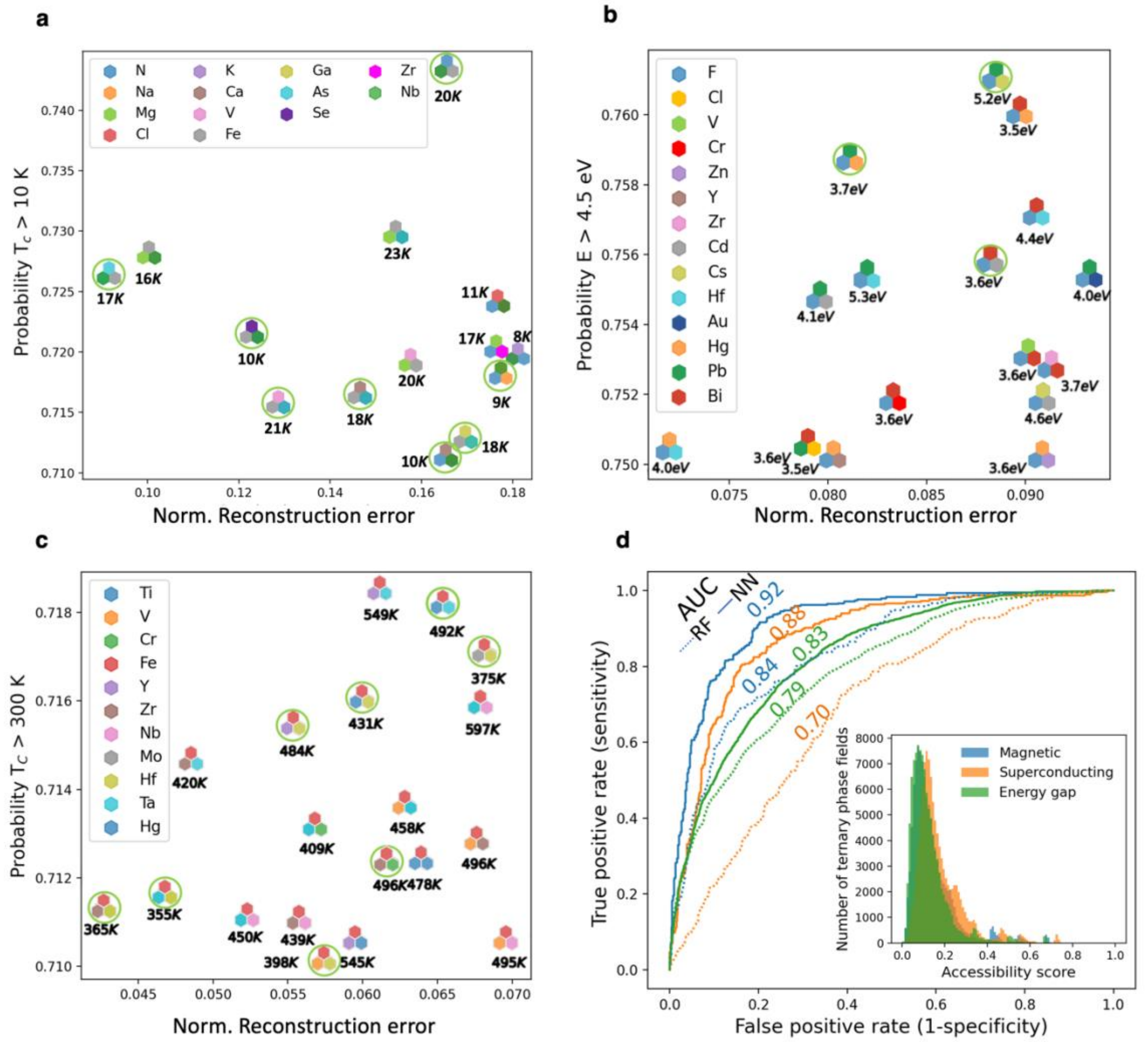

**Figure 5. Probability of high-value properties, prediction of maximum property values and normalised reconstruction errors as a measure of similarity with synthetically accessible materials for unexplored ternary phase fields. Materials reported in ICSD[2], for which property values are not in SuperCon-v2018[3,24] or MPDS[1] are circled. A** Unexplored ternary phase fields that are classified to exhibit

superconductivity at T > 10 K with more than 70% probability and that have high likelihood of forming stable compounds with normalised reconstruction error < 0.2, demonstrate trends in constituent elements: most of the top 50 phase fields are predicted to contain Mg, Fe, Nb and N. **b** Unexplored ternary phase fields that are classified to exhibit an energy band gap > 4.5 eV with more than 75% probability and that have high likelihood of forming stable compounds (with reconstruction error < 0.1) demonstrate trends in distribution by constituent elements: different combinations of Hg-, F-, Bi-, Hf- and Pb-based phase fields have the highest probabilities. **c** Unexplored ternary phase fields that are classified to exhibit magnetic properties at Curie T > 300 K with more than 71% probability and that have high likelihood of forming stable compounds (with reconstruction error < 0.1) demonstrate trends in constituent elements: all top-ranked phase fields are Fe-based, with many phase fields containing Co and Y. **d** Receiver operating characteristics (ROC) of the classification models trained and tested on the random 80-20% train-test splits demonstrate high sensitivity and specificity of classifications for the range of thresholds of probabilities. The corresponding areas under the curves (AUC) demonstrate overall excellent performance of the model for magnetic materials, and good performance for both superconducting transition temperature and energy gap classifications. PhaseSelect considerably increases AUC for all datasets in comparison to Random Forest classifiers trained on the same data. The inset illustrates close match of the distributions of 105995 ternary phase fields with respect to their reconstruction errors for all three datasets.

By applying PhaseSelect to 105995 ternary phase fields and focusing on the 90029 unexplored phase fields (Supplementary discussion) that do not have any related compositions with reported properties in MPDS or SuperCon-v2018, we classify new elemental combinations with respect to the threshold values of superconducting transition temperature, Curie temperature and energy band gap, predict maximum values for these properties expected within the phase fields, and orthogonally rank candidate phase fields by their reconstruction errors – degree of similarity with experimentally synthesized materials that are reported to exhibit these properties. We also highlight the phase fields, where compositions were synthesized and reported in ICSD, but for which there is no information about the properties discussed here in SuperCon or MPDS (encircled markers in Figure 5a-c), hence these phases fields did not enter the data for training. The large number of such phase fields among the top-performing candidates provides verification of the developed models and demonstrates that highly ranked candidates are likely to produce thermodynamically stable materials observed experimentally. We report the full list of likely candidates for novel superconducting materials among the phase fields that have been reported to form stable compounds in ICSD, but were not investigated from the perspectives of superconducting applications (See the full list of candidates in [30] and its excerpt in Supplementary Table 7).

The top-performing phase fields according to both the probability of exhibiting high values of properties and conformity with synthetically accessible materials demonstrate trends produced by the constituent chemical elements: Mg, Fe, Nb are predicted to constitute most of the top 50 phase fields that would yield stable compositions with superconducting transition temperatures above 10 K; similarly the top 50 magnetic ternary materials are Fe-based; while different combinations of Bi, Hf, Hg, Pb and F are predicted as most likely phase fields to contain stable compounds with energy gap of more than 4.5 eV, which can be expected from simple bonding considerations as the majority of the latter are fluorides.

While these predictions may align well with the human experts' understanding of chemistry, hence emphasizing the models' capability to infer complex elemental characteristics in relation to properties from historical data, the models can also be used to identify unconventional and rare prospective elemental combinations as well as to rank the attractive candidate materials for experimental investigations. Such less expected examples include combinations of elements that do not exhibit ambient pressure or pressure-induced superconductivity as elemental solids[31], exclude Fe, Cu and rare-earth metals, known for forming families of superconducting materials, but are classified as high-temperature (>10K) superconductors, when combined: C-Mg-Rb, Cr-K-N and As-C-Na among other 125 ternaries[30]. For magnetic applications, all unexplored phase fields that were classified to exhibit magnetism at $T_c$ > 300K contain known magnetic elements Fe, Co, Ni, or Mn. However, we can highlight the combinations that are interesting also from the perspective of similarity to the synthetically accessible materials, including 4001 Fe-free ternaries[30], such as Co-Ti-Zr, Mo-Co-B and Hf-Mn-Ti. Among 13070 large bandgap dielectric phases[30] not involving oxides nor fluorides we can highlight Te-S-I and Ga-S-Cl.

## Conclusions

The selection of elements as material components is the cornerstone of materials design, as their choice delimits all future outcomes in subsequent synthetic work. Quantitative assessment of the potential properties of the prospective materials at the level of their constituent elements mitigates the high risk of the consequential decisions in elaborate research of materials discovery. Classification of the materials for functional applications agglomerated into phase fields avoids the challenges of the common composition-based approaches, such as the exhaustive assessment of all possible combinations and in-cluster extrapolation without data leakage. Working at the level of phase fields is also a route to reducing the combinatorial space by several orders of magnitude.

The end-to-end integrated architecture of PhaseSelect represents and quantifies phase fields in two orthogonal dimensions: functional performance (via classification and regression), and similarity to synthetically accessible materials, which can be coupled or used independently for any combination of elements at scale to prioritise experimental targets. By employing PhaseSelect at the conceptualization stage of material discovery and synthesis, human researchers can make use of numerical guidance in the selection of elements that are most likely to produce new stable compounds with a high probability of superior functional properties. This enables the combination of statistically derived quantitative information with the expert knowledge and understanding to prioritise promising phase fields and de-risk material discovery. Finally, the attention mechanism of PhaseSelect presents a route to interpretation of the machine learning for materials science and allows extrapolation of the knowledge of materials databases to the large number of unexplored phase fields. These include multi-element materials, with prospective performance that could not otherwise be computationally assessed at scale with current methods.

## Methods

In this work, we adopt the same architecture for all three problems investigated.

For unsupervised learning of elemental representations the shallow autoencoder has a single hidden layer with ReLU activation[26], and sigmoid activation for the decoder layer. Effects of different numbers of nodes on the model performance are studied (Supplementary Fig. 1).

We employ 8-head attention[20] for learning the weights for elemental contributions in the phase fields representation and a padding mask $\mu$ for length justification of phase fields with different number of chemical elements by a maximum-size phase field: $\mu$: $\mathbb{R}^{k\cdot m} \to \mathbb{R}^{k\cdot \max(m)}$, $\mu(\boldsymbol{p}_{k_1 \dots k_m}) = \boldsymbol{p}_{k_1 \dots k_m z_{m+1} \dots z_{m+\max(m)}}$, $z_j = 0$. Multi-head approach ensures stabilisation of the training and improvement of the performance. For classification neural network we use 2 hidden layers with 80, 20 nodes respectively, with ReLU activation, L1 = 0.03, L2 = 1e-4 regularizations and 0.5 dropout[27].

The ranking AutoEncoder is built with 4 hidden layers for encoder with decreasing number of nodes, 1/2, 1/4, 1/8 and 1/16 respectively of the initial length of a phase field vector, 4-dimensional latent representation and 4 hidden layers in decoder with increasing number of nodes, 1/16, 1/8, 1/4 and 1/2 of the initial length of a phase field vector. Each AutoEncoder hidden layer is followed by 0.1 dropout and activated with ReLU.

For the training, we employ Adam optimisation[32] with a starting learning rate of 1e-3 and a scheduled decrease after each 100 epochs. During training, we monitor the accuracy (mean absolute error for regression) and ensure early stopping with 7 (40) epoch patience respectively.

## Acknowledgements

We thank the UK Engineering and Physical Sciences Research Council (EPSRC) for funding through grants number EP/N004884 and EP/V026887. D. A, V. G., and M.W.G. thank the Leverhulme Trust for funding via the Leverhulme Research Centre for Functional Materials Design.

## Data availability

The raw data used in this study is available at https://www.github.com/lrcfmd/PhaseSelect. The distribution of the phase fields' rankings, computed phase field's probability data generated in this study are available via University of Liverpool data repository at https://doi.org/10.17638/datacat.liverpool.ac.uk/1613

## Software availability

The software developed for this study is available at https://www.github.com/lrcfmd/PhaseSelect and at DOI:10.5281/zenodo.7464312.

## Author Contributions

A.V. identified, developed, and implemented the PhaseSelect model in discussion with D.A., V.G., M.W.G., and M.S.D. A.V., D.A., and M.J.R. wrote the first draft, all authors contributed to the completion of the manuscript. M.J.R. directed the project.

## Competing Interests Statement

The authors declare there are no competing interests.

## Supporting Information

Machine Learning methodology and models, Training set for the Classification, Model validation, Elemental attention scores, Tools and libraries.

# Supplementary Information

# Element selection for functional materials discovery by integrated machine learning of atomic contributions to properties

Andrij Vasylenko[1], D. Antypov[1], V. Gusev[1], M. W. Gaultois[1], M. Dyer[1], M. J. Rosseinsky[1,*]

[1]Department of Chemistry, University of Liverpool, L697ZD Liverpool, UK

[*]corresponding author

## Contents

## Atomic features encoding

For unsupervised learning of atomic features from the materials database[1], we employ an approach similar to reference[2], in which we substitute single value decomposition (which may be inadequate for processing binary data of size (n, d), where n = 87 is a number of chemical elements, d = 212991 is a number of compositions considered, d >> n)[3,4] with a shallow autoencoder. A shallow autoencoder is a 3-layer neural network, in which the input and output layers have a large number of neurons that corresponds to the size of the input vectors – sparse binary encoding representations of atoms in the database. A single latent layer in between the input and output is a bottleneck aiming to extract the essential patterns in the data, while decreasing its dimensionality and filtering out the less representative and noisy information. One can further use thus trained representations as the atomic features. To maximise the quality and the descriptive power of the extracted atomic features, we study the effect of the size of the latent layer on the metrics of the downstream classifications. In this work, we train the shallow autoencoder simultaneously with the classification neural network in the end-to-end fashion. When trained separately for classification of superconducting materials, magnetic materials, and materials with a reported band gap, the end-to-end models based on the different sizes of atomic vectors have the metrics depicted in Supplementary Figure 1 a, b, c respectively. Although the best performance for classification of different properties is achieved at different numbers of atomic features in each of the three cases, there is similar trend for these dependencies. This trend suggests that a small number (< 40) of features cannot fully capture the variation in data, and a large set of features (> 80) contains too much noise, hence there is an optimal number of atomic descriptors for each model.

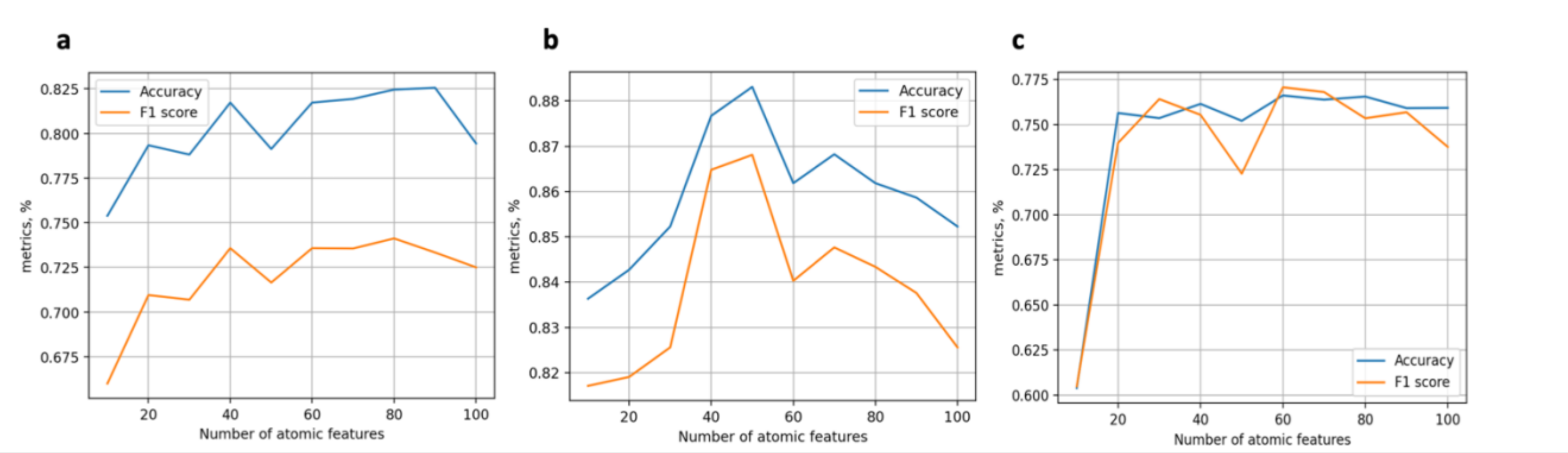

**Supplementary Figure 1. Changes in classification metrics for models with different numbers of atomic features. a** Accuracy and F1 score for classification of materials with respect to the maximum of superconducting transition temperature (classes are divided on above and below $T_c$ =10K): the best performing model has 80 atomic features; **b** Accuracy and F1 score for classification of materials with respect to the maximum of Curie transition temperature (classes are divided on above and below $T_c$ = 300 K): the best performing model has 50 atomic features; **c** Accuracy and F1 score for classification of materials with respect to the maximum of energy bandgap (classes are divided on above and below $E_g$ = 4.5 eV): the best performing model has 60 atomic features.

### Attention to atomic contributions maximizing the properties

In the end-to-end classification models, we employ an attention mechanism[5] to emphasize those atomic contributions that minimize the combined loss, and hence maximize classification metrics.

To incorporate information about atomic bonding interplay from all available data, the variance in size of the phase fields is alleviated by zero-padding in the phase fields representation module that further allows extrapolation of the patterns derived from the explored materials onto the candidate phase fields of arbitrary number of elements. We extract the attention scores obtained during the training of the models that illustrate atomic contributions to the properties manifested by the phase fields (Supplementary Figures 2-6). For visualisation, the attention scores are averaged across the attention heads and across all instances of the atomic pairs in the corresponding datasets. In Supplementary Figure 11, distributions of the averaged attention scores are plotted for the atoms that contribute the most to identify phase fields that manifest particular properties.

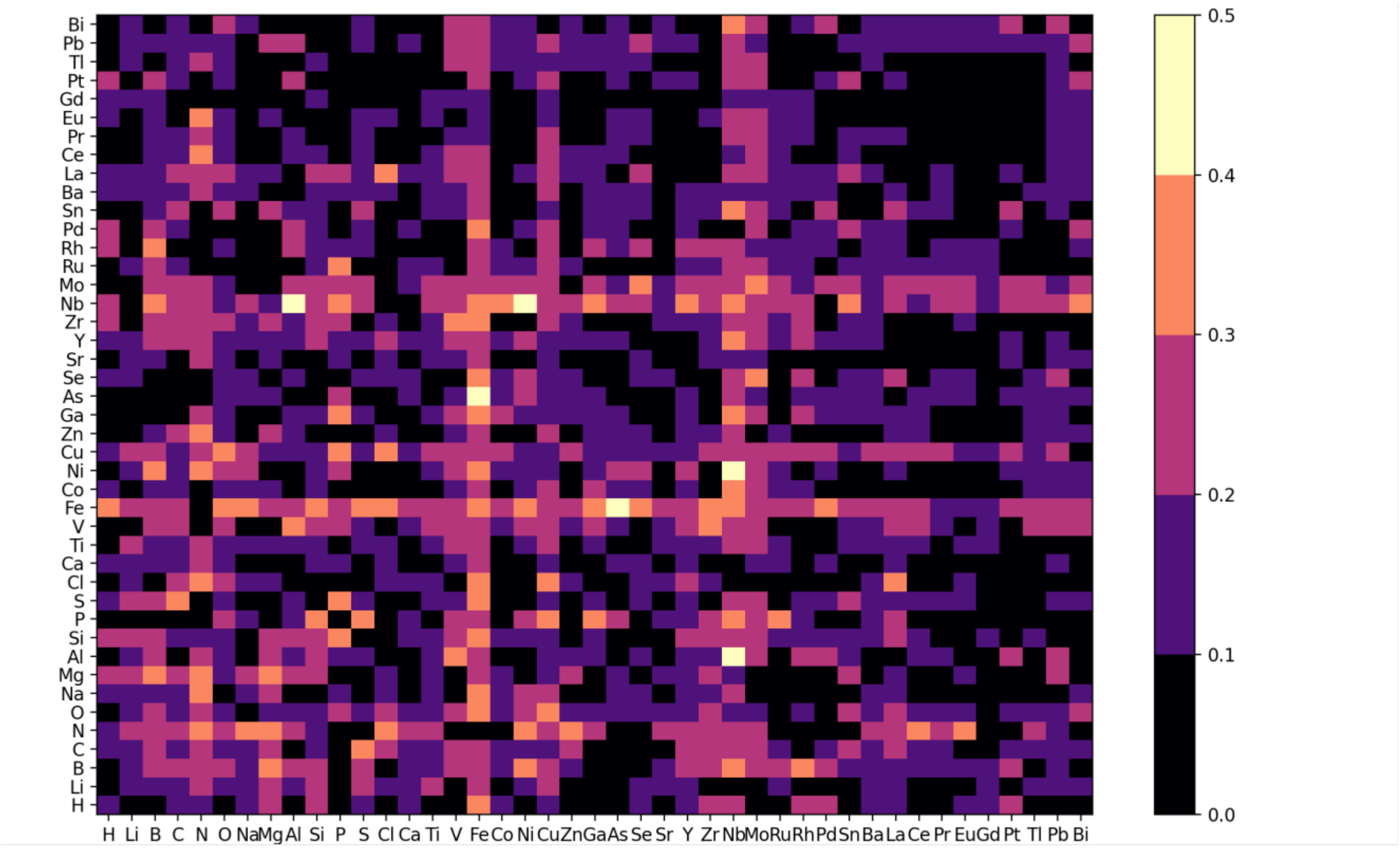

**Supplementary Figure 2. Attention to atomic pairs that maximize accuracy of classification of high-temperature superconducting materials.** Attention scores vary from 0 to 1. By focusing on the atomic pairs with the highest scores, when describing the phase field, accuracy of classification of these phase fields is maximized. This suggests the atomic pairs with the most prominent contributions allowing high-temperature superconductivity, e.g. Nb-Al, Nb-Ni, Cu-O and Fe-As.

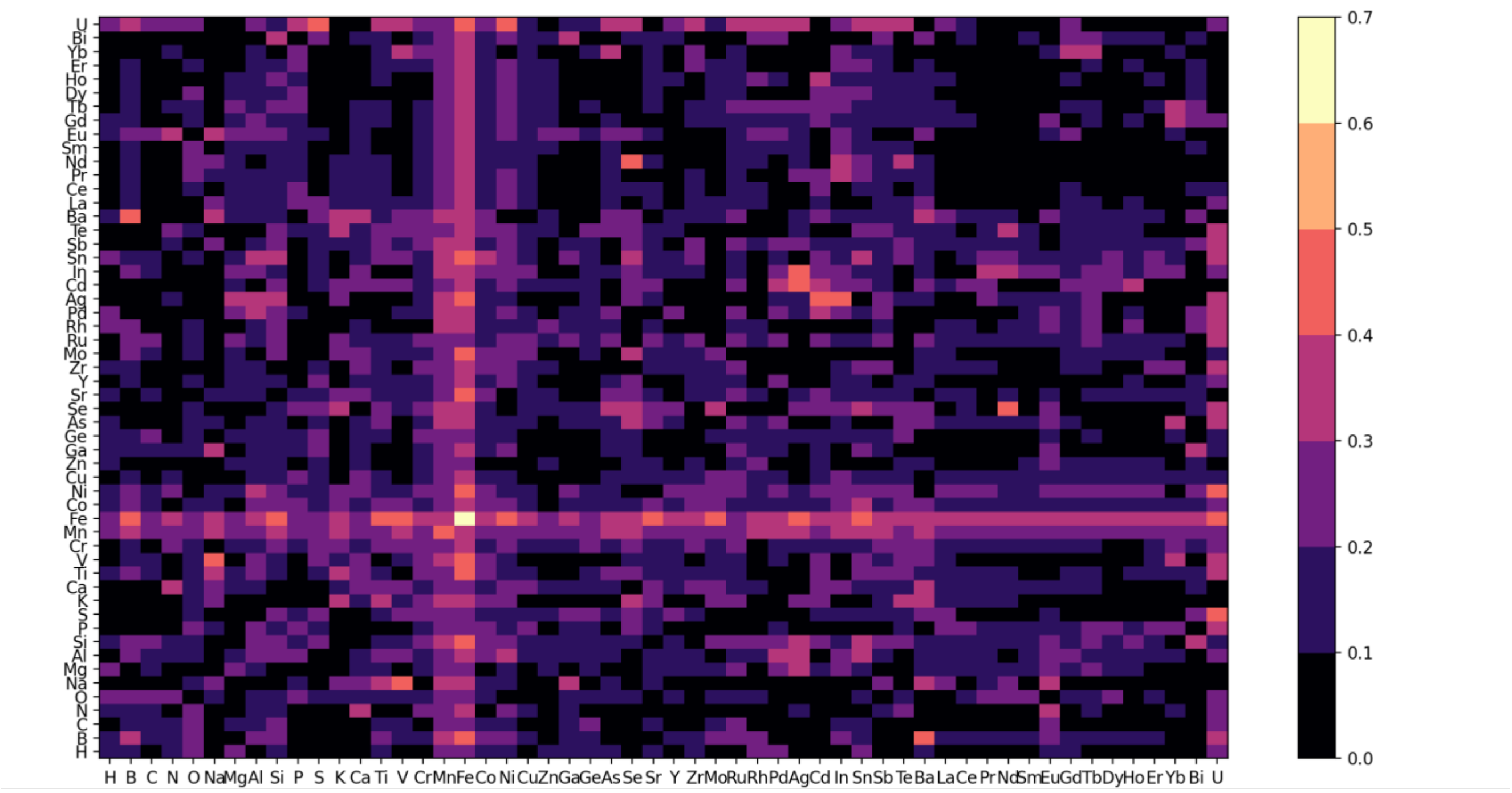

**Supplementary Figure 3. Attention to atomic pairs that maximize the accuracy of classification of high-temperature magnetic materials.** Attention scores vary from 0 to 1. By focusing on the atomic pairs with the highest scores, when describing the phase field, accuracy of classification of these phase fields is maximized. This suggests the atomic pairs with the most prominent contributions allowing high-temperature magnetic behaviour, with Mn, Fe, Ni and Co included in the majority of such pairs.

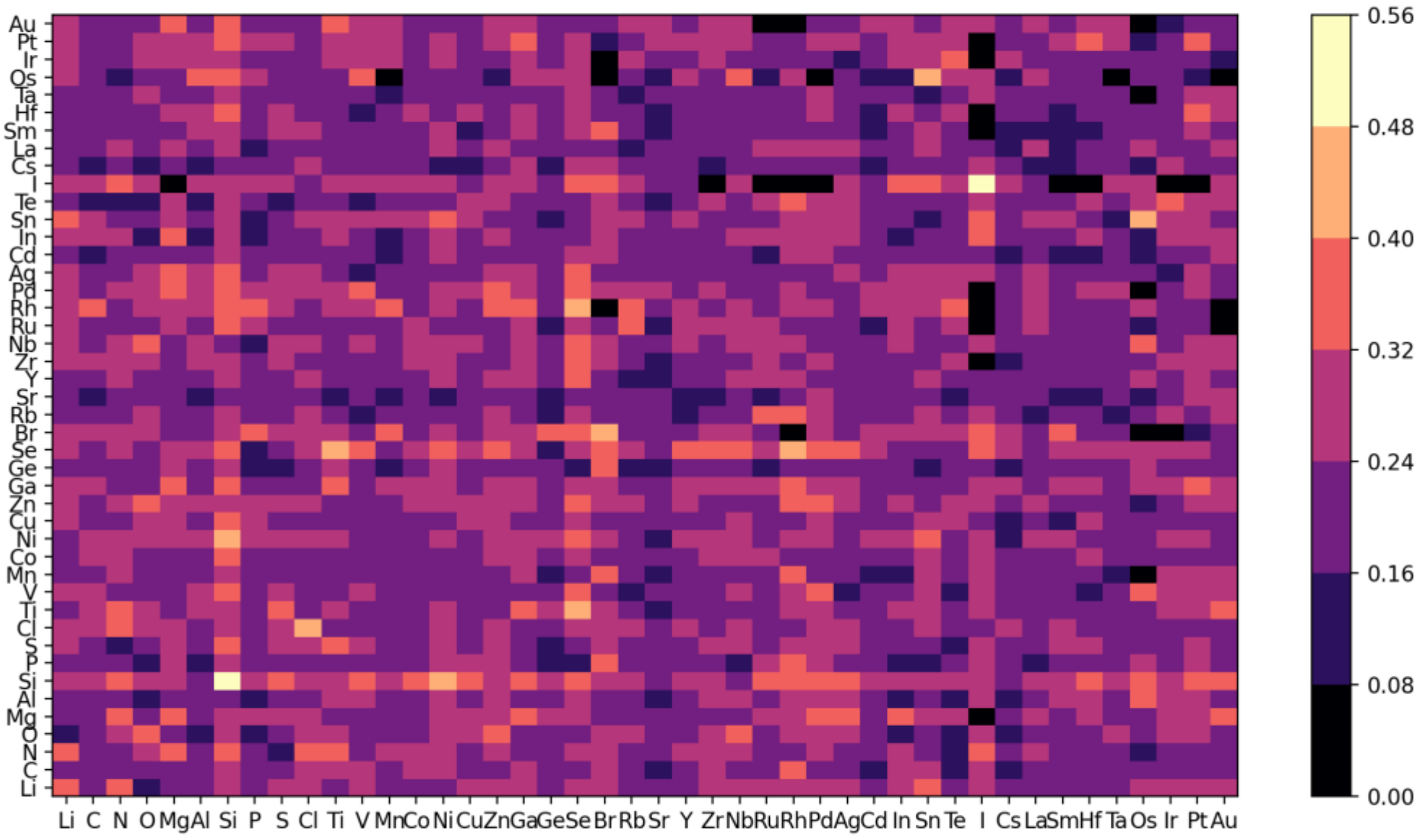

**Supplementary Figure 4. Attention to atomic pairs that maximize accuracy of classification of materials with energy gap <4.5 eV.** Attention scores vary from 0 to 1. By focusing on the atomic pairs with the highest scores, when describing the phase field, accuracy of classification of these phase fields is maximized. The majority of the atoms in the phase fields have 0.3-0.5 attention score, and contribute equally to identification of low energy gaps.

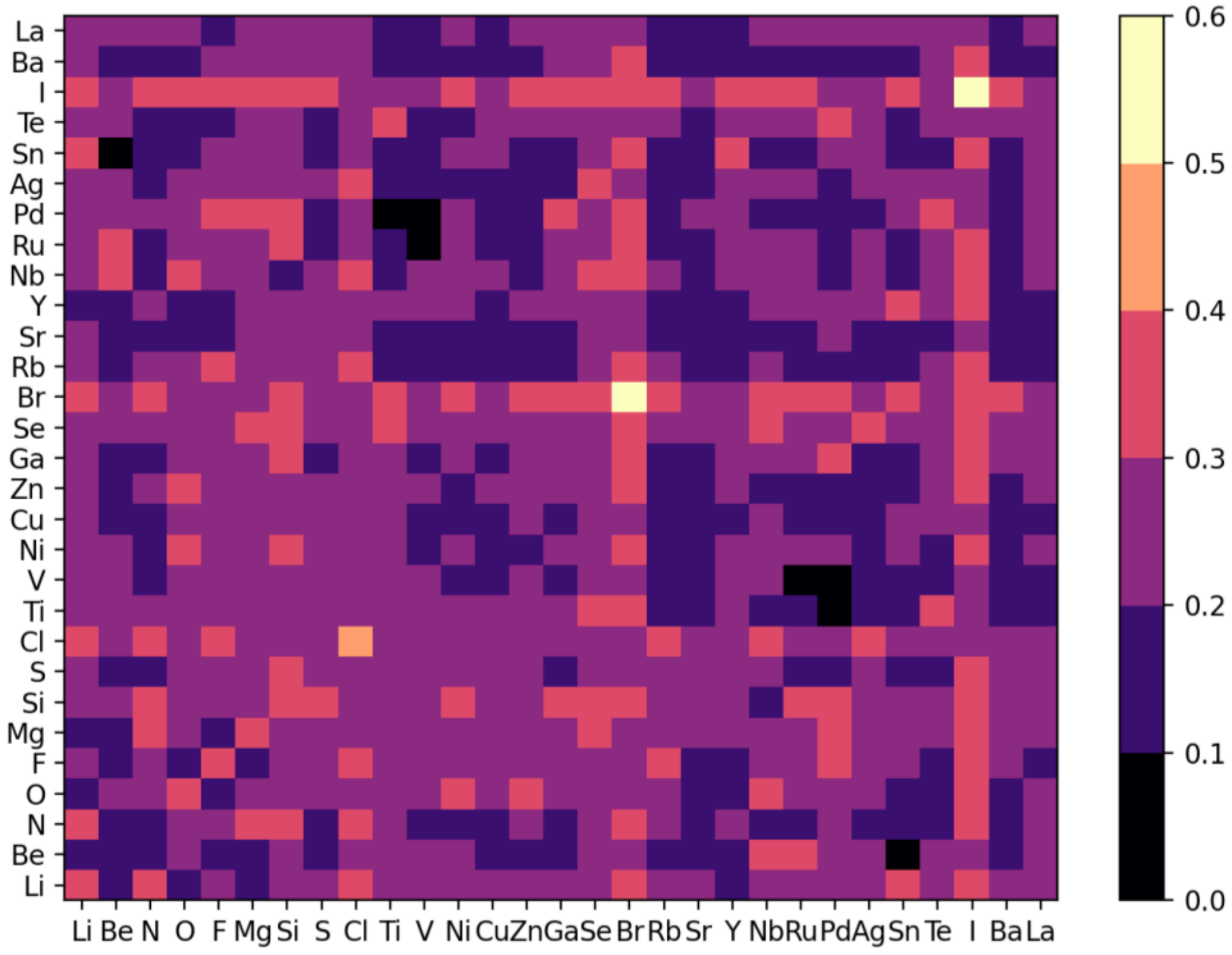

**Supplementary Figure 5. Attention to atomic pairs that maximize accuracy of classification of materials with energy gap > 4.5 eV.** Attention scores vary from 0 to 1. By focusing on the atomic pairs with the highest scores, when describing the phase field, accuracy of classification of these phase fields is maximized. This suggests atoms and atomic pairs with the most prominent contributions to the materials with energy gap > 4.5 eV, e.g. I, Br, Se, Cl, Si.

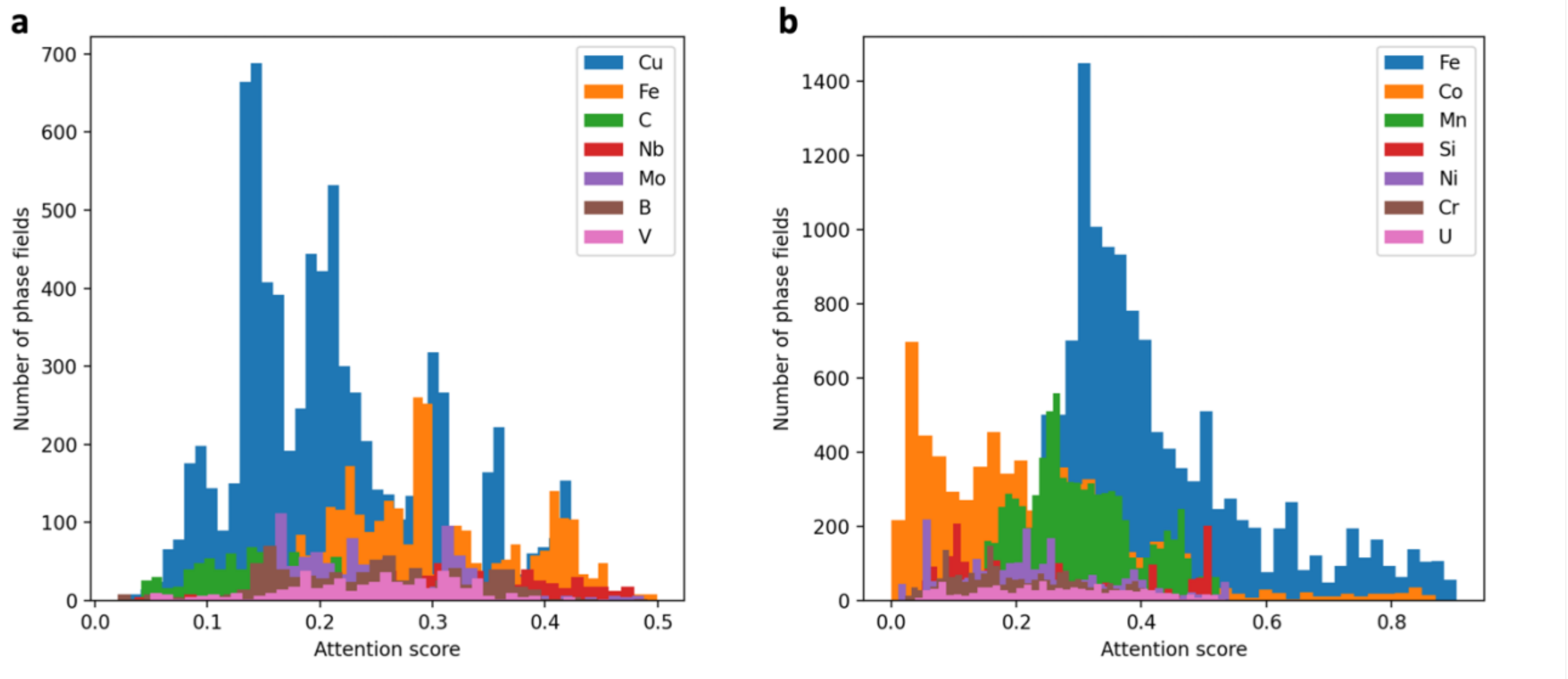

**Supplementary Figure 6. Distribution of attention scores for the most contributing atoms to the functional materials a** High-temperature superconducting materials; **b** high-temperature magnetic materials.

The atomic contributions weights are also used for building a model for an arbitrary number of elements in a phase field. For this, we create all phase fields representations vectors of an equal size *l*, corresponding to the largest phase field in a database, and pad the smaller phase field vectors, of size *s*, with *l* - *s* zeros, that will have zero attention weights, but will further enable formation of a neural network layer for processing of all input data in a single model. The described construction of a phase field representation with local attention weights also makes the model insensitive to the order in which atomic elements are listed in a phase field, without the need to take into account all possible permutation of the elements.

## Supervised model training and validation

To validate model performance, we employ 5-fold cross-validation for each dataset: phase fields with reported values of superconducting transition temperature, phase fields with reported values of Curie transition temperature, phase fields with reported values of energy gap as illustrated in a diagram in Supplementary Figure 7a.

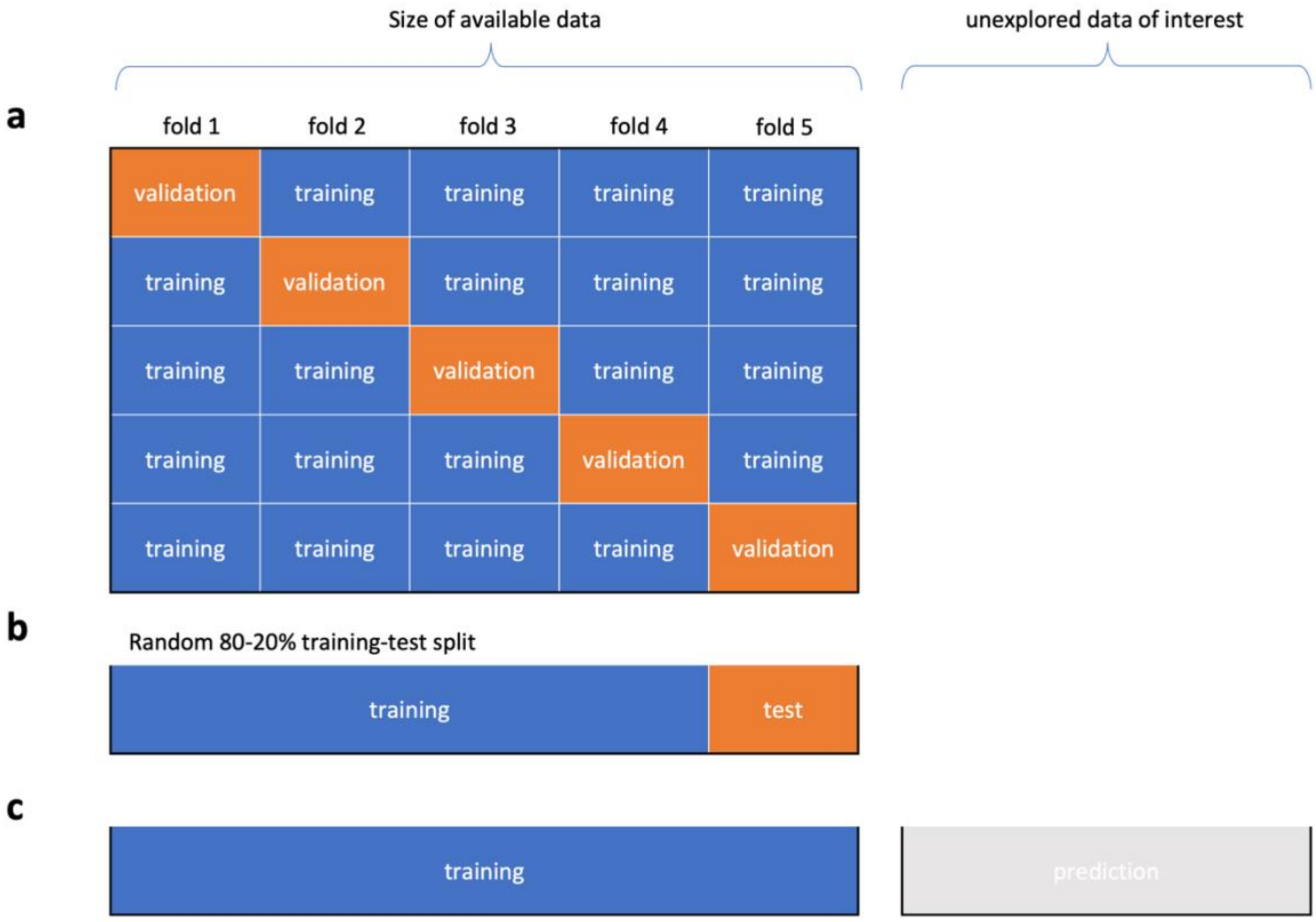

**Supplementary Figure 7. Datasets division for validation, illustration and prediction. a** dataset non-overlapping splits for 5-fold cross-validation[6] and benchmarking[7]; **b** random 80-20% training-test data split for model performance illustration; **c** full data is used for training in a production mode for prediction of properties of unexplored phase fields.

In 5-fold cross-validation, the data is divided into the training and validation sets (80% and 20% of data respectively) in 5 different non-overlapping subsets so models are examined with respect to the ability of the chosen architecture to generalise and extrapolate the information learnt from 5 different subsets of data (presumably presenting different chemistry of phase fields and property values) onto the unseen areas. MAE for regressors, accuracy and F1 scores of the classification models are presented in Supplementary Table 1.

**Supplementary Table 1. PhaseSelect MAE for regression and accuracy and F1 scores for classification models in 5-fold cross-validation**

| | Superconducting $T_c$=10K | | Magnetic $T_c$=300K | | Energy gap 4.5 eV | |
|---|---|---|---|---|---|---|
| **Regression** test data subset | MAE, K | MAE / $T_c$ Range, % | MAE, K | MAE / $T_c$ Range, % | MAE, eV | MAE / Gap Range, % |
| 0-20% | 10.4 | 7.2 | 120.1 | 5.5 | 42.8 | 6.1 |
| 21-40% | 8.2 | 5.7 | 119.4 | 5.4 | 44.4 | 6.3 |
| 41-60% | 7.4 | 5.1 | 125.4 | 5.7 | 44.3 | 6.3 |
| 61-80% | 7.0 | 4.9 | 133.0 | 6.1 | 44.8 | 6.4 |
| 81-100% | 7.2 | 5.0 | 137.8 | 6.3 | 44.2 | 6.3 |
| **Average:** | **8.0** | **5.6** | **127.0** | **5.8** | **44.1** | **6.3** |
| **Classification** test data subset | Accuracy,% | F1 score,% | Accuracy,% | F1 score,% | Accuracy,% | F1 score,% |
| 0-20% | 80.9 | 73.3 | 86.8 | 84.5 | 75.5 | 75.6 |
| 21-40% | 83.6 | 77.1 | 86.7 | 85.7 | 75.2 | 74.8 |
| 41-60% | 78.7 | 71.7 | 85.9 | 82.1 | 75.9 | 75.6 |
| 61-80% | 79.7 | 71.3 | 85.5 | 84.1 | 76.0 | 75.1 |
| 81-100% | 79.2 | 71.0 | 86.0 | 84.4 | 75.7 | 75.5 |
| **Average:** | **80.4** | **72.9** | **86.2** | **84.2** | **75.6** | **75.3** |

The performance metrics from the 5 models for each dataset are then averaged to describe a general ability of the models' architecture to learn from the available data. During the training of the end-to-end classification models, the weights and biases of the autoencoder for atomic features extraction and classifier neural networks are trained simultaneously, while the corresponding losses – reconstruction error and binary cross-entropy, respectively – are minimized as a combined loss during back propagation with Adam optimization[8]. The typical training of the regression and classification models for the superconducting, magnetic and energy band gap datasets are illustrated in the Supplementary Figure 8 and 9 respectively.

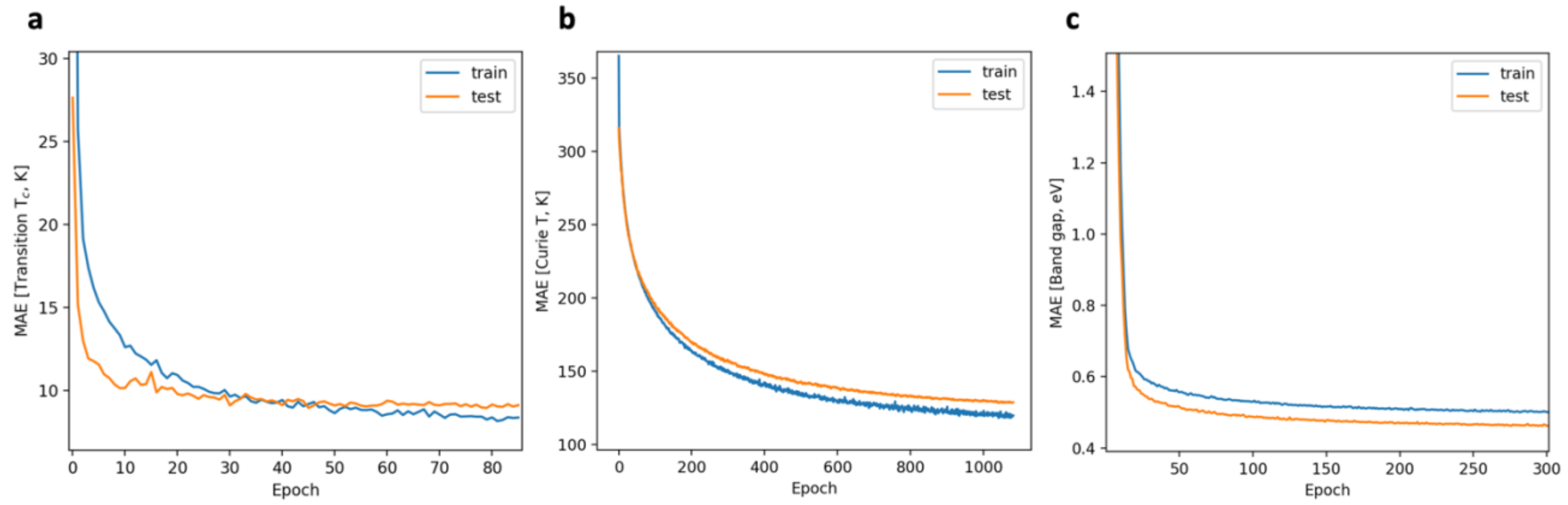

**Supplementary Figure 8. Training progress of the end-to-end regression models. a** Regression of the superconducting materials, training on 4826 phase fields; MAE loss in K; **b** Regression of the magnetic materials, training on 4753 phase fields; MAE loss in K; **c** Classification of the materials' energy band gap, training on 40452 phase fields; MAE loss in eV.

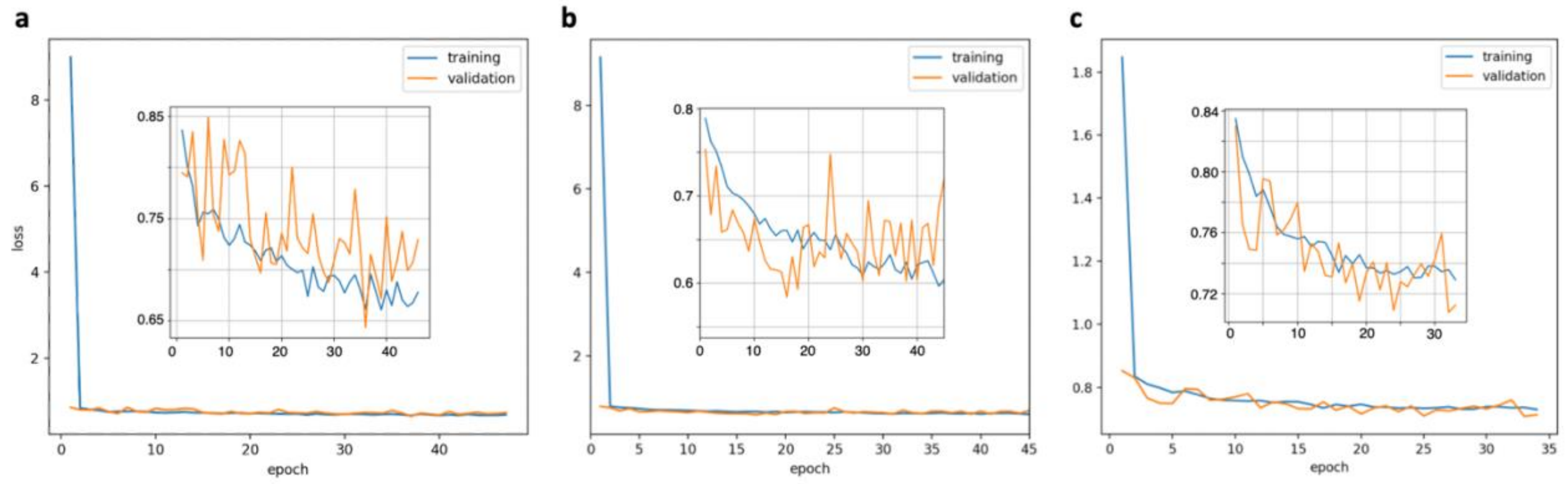

**Supplementary Figure 9. Training progress of the end-to-end classification models. a** Classification of the superconducting materials, training on 4826 phase fields; **b** Classification of the magnetic materials, training on 4753 phase fields; **c** Classification of the materials' energy band gap, training on 40452 phase fields.

## Benchmarking against Random Forest and dummy regression

In comparison with dummy regression, where each phase field is assigned with a mean value in a dataset, random forest regressors[6] trained separately for the studied datasets demonstrate an improved performance (Supplementary Fig. 10).

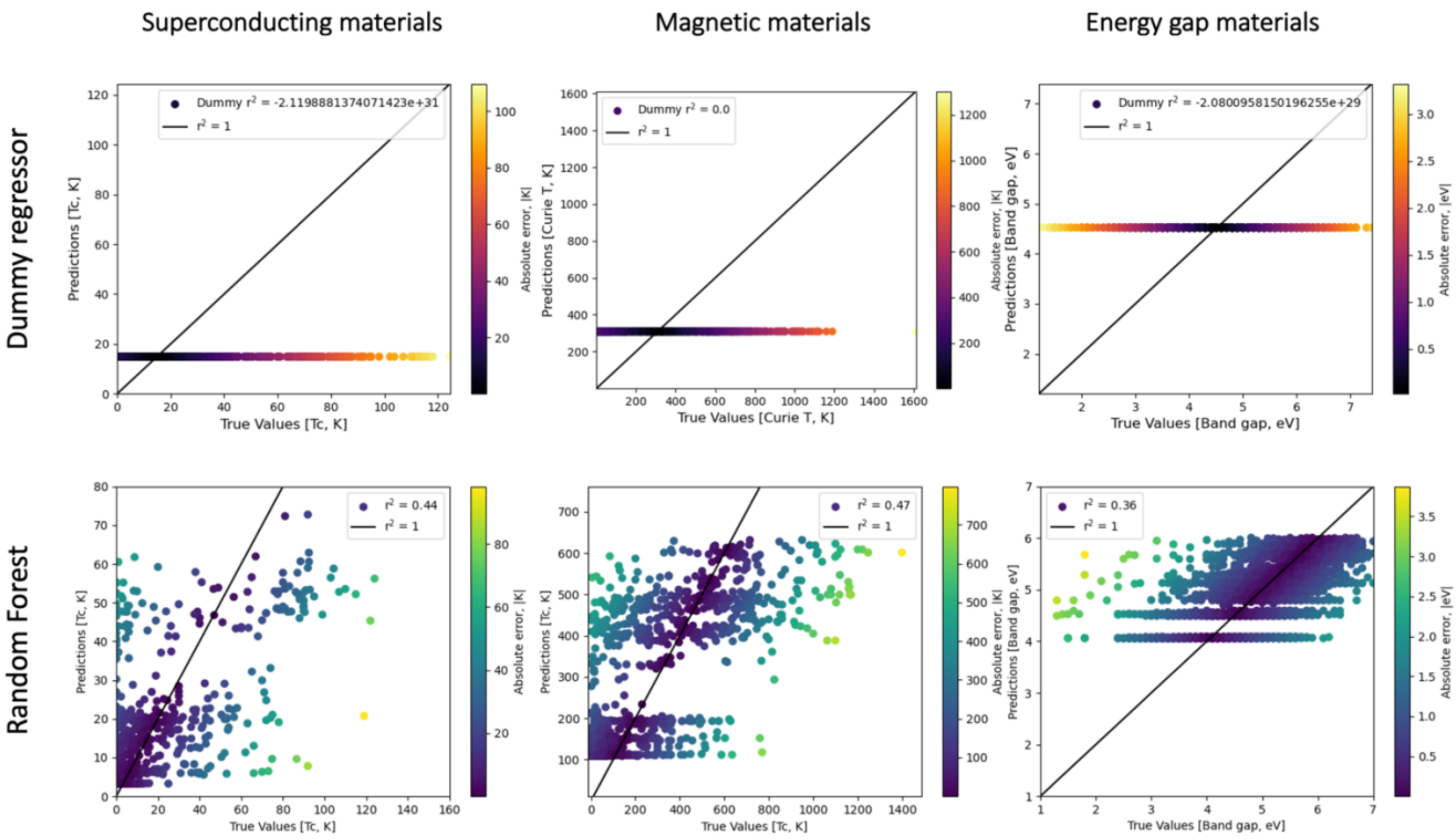

**Supplementary Figure 10. Dummy regression and random forest regressors for superconducting, magnetic materials and materials with reported energy gap.** Random forest models are trained separately for 3 datasets. The performance is illustrated on random 80-20% train-test splits, and is characteristic to the metrics observed in k-fold cross-validation (Figure 4 in main text).

## Unsupervised model training and validation

For validation of the unsupervised models for the phase fields ranking with respect to synthetic accessibility, we employ an approach developed in [9]. We perform 5-fold cross-validation, in which the validation error is defined as the percentage of entries in the test set that evaluated with normalized reconstruction errors in the 20% of the maximum Supplementary Table 2. Additionally, we compare the predicted reconstruction errors for the validation sets with the ground truth reconstruction errors obtained for the same entries in unsupervised training, when the entries are included in the training data (Supplementary Fig. 11) and calculate the mutual information score adjusted against chance[10] (Supplementary Table 2). The typical training process of the ranking autoencoder neural network for different datasets are depicted in Supplementary Fig. 12.

**Supplementary Table 2. Accuracy and Adjusted Mutual Information Score (AMIS) for ranking autoencoder models in 5-fold cross-validation**

| | **Superconducting materials** | | **Magnetic materials** | | **Energy gap materials** | |
|---|---|---|---|---|---|---|
| test data subset | Accuracy,% | AMIS | Accuracy,% | AMIS | Accuracy,% | AMIS |
| 0-20% | 96.1 | 0.69 | 94.7 | 0.64 | 97.2 | 0.77 |
| 21-40% | 97.4 | 0.76 | 95.3 | 0.66 | 98.6 | 0.78 |
| 41-60% | 97.7 | 0.68 | 93.5 | 0.66 | 97.7 | 0.75 |
| 61-80% | 95.1 | 0.79 | 94.9 | 0.64 | 98.7 | 0.75 |
| 81-100% | 96.6 | 0.72 | 93.9 | 0.68 | 97.8 | 0.81 |
| **Average:** | **96.6** | **0.73** | **94.5** | **0.66** | **98.0** | **0.77** |

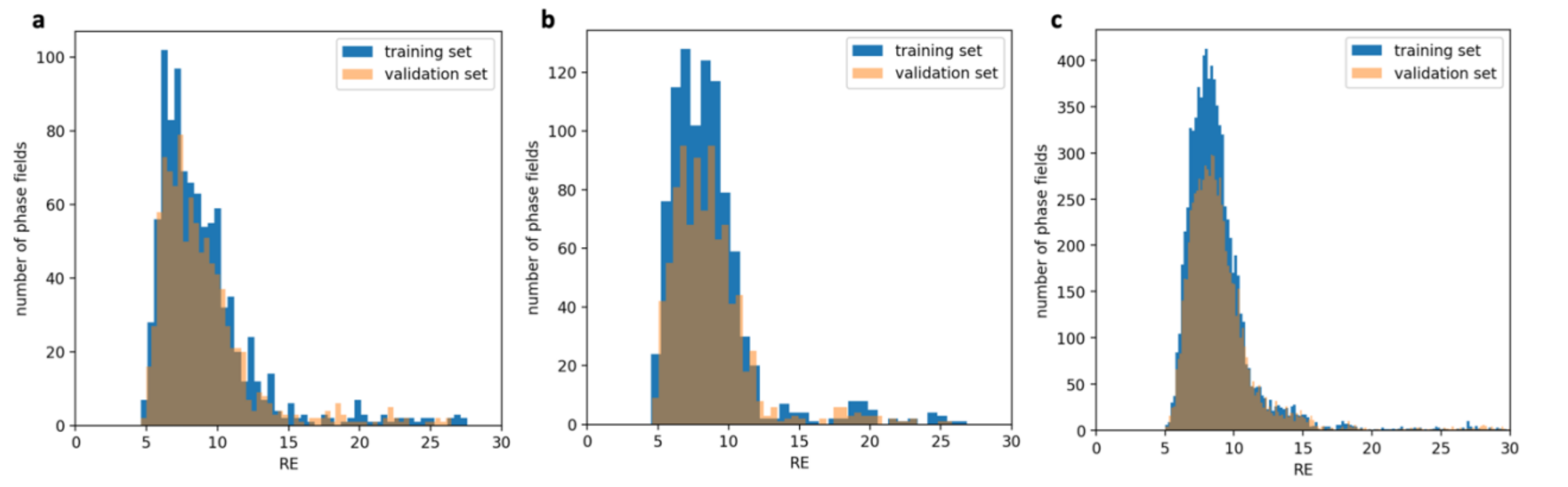

**Supplementary Figure 11. Distribution of reconstructions errors (RE) for the phase fields.** RE for the same phase fields are calculated in two approaches: 1) in unsupervised learning, as a part of a training set – used as ground truth RE for AMIS calculation in Supplementary Table 2; 2) predicted by the model trained on 80% of the remaining data – as a validation set. **a** Superconducting materials; **b** magnetic materials; **c** materials with reported energy gap.

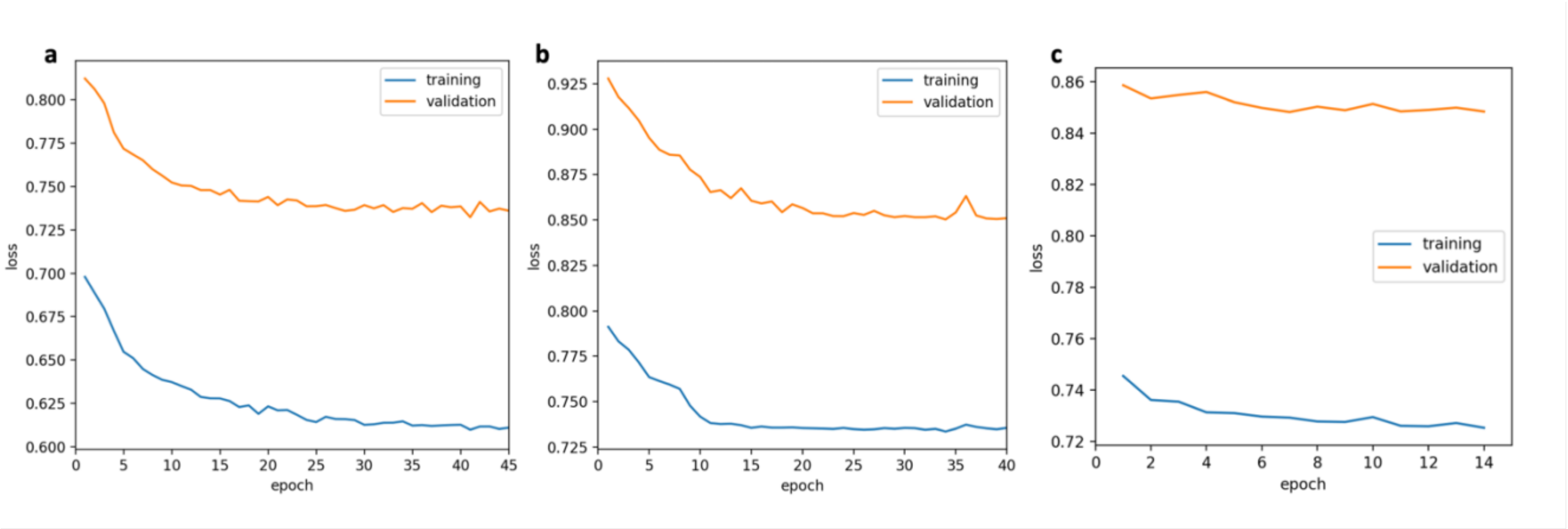

**Supplementary Figure 12. Training progress of the ranking autoencoder models. a** ranking of the superconducting materials, training on 4826 phase fields; **b** ranking of the magnetic materials, training on 4753 phase fields; **c** ranking of the materials with the reported energy band gap, training on 40452 phase fields.

To take into account statistical variance in both supervised and unsupervised results from the neural networks trained at different instances, we average the results across the ensemble of 250 neural networks. Convergence of deviations of results in terms of the mean square errors from the running average values is illustrated in Supplementary Figure 10. For all datasets, for both supervised classifying neural network and ranking autoencoders, the average values converge when more than 200 models are considered.

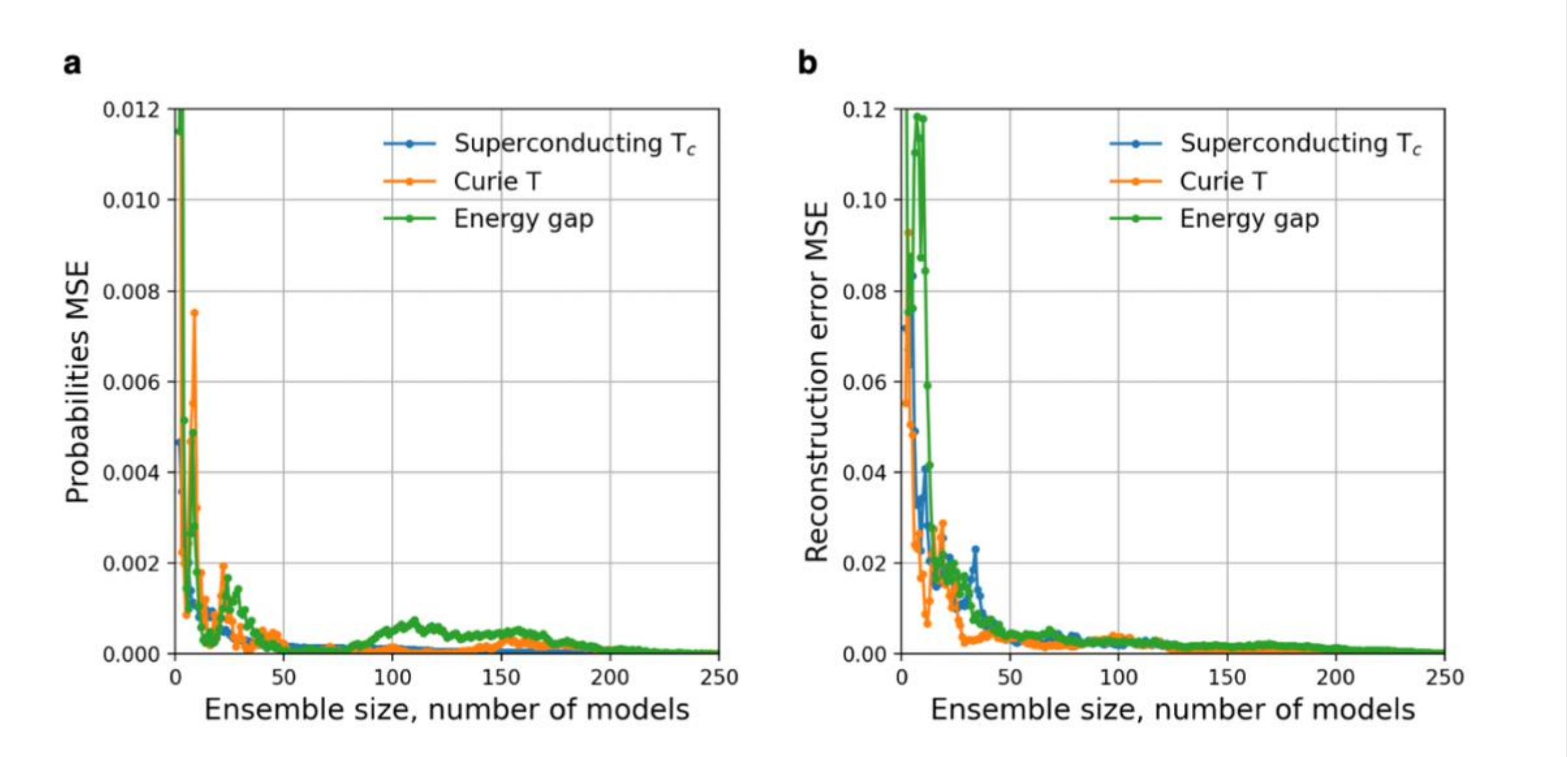

**Supplementary Figure 13. Convergence of the mean square errors (MSE) of the average predicted scores with the number of models in the ensemble. a** Probabilities of phase fields belonging to a binary class are averaged over an ensemble of models. MSE of the average scores decrease below 0.001 for ensembles larger

than 200 models for all datasets. **b** MSE of the average reconstruction errors, used as synthetic accessibility scores of phase fields decrease below 0.005 for ensembles larger than 200 models.

The ensembles of the trained models for each dataset are then used to classify the phase fields with respect to the corresponding properties. For randomly selected 20% of the phase fields from each dataset, the classification predictions are illustrated with the confusion matrices in Supplementary Figure 11.

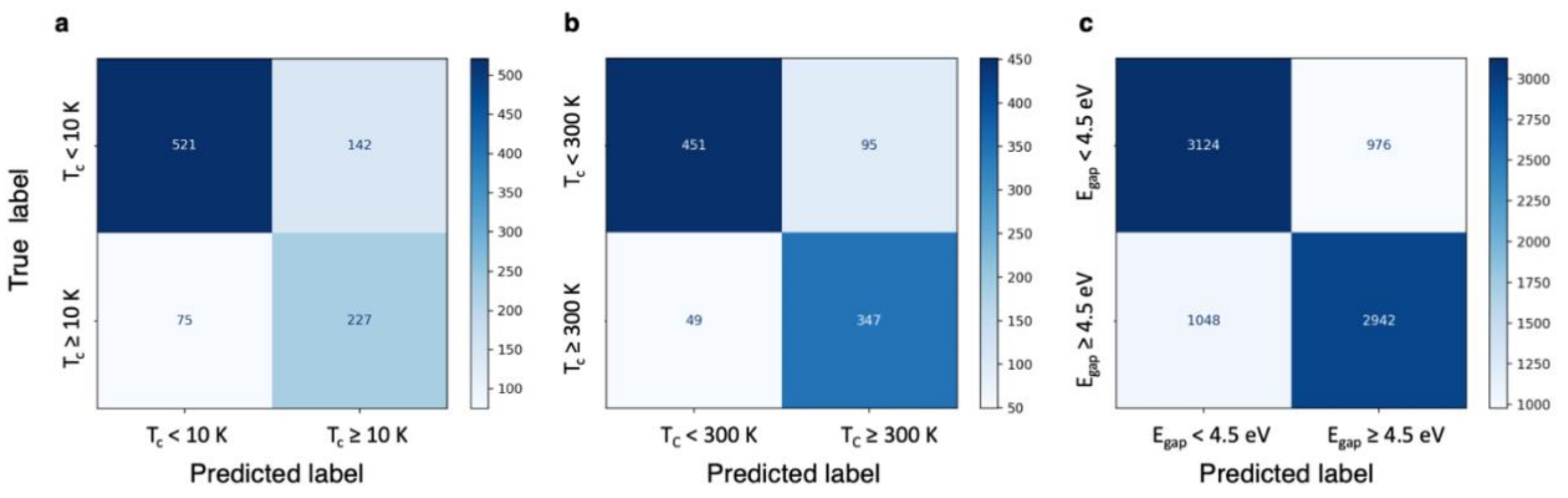

**Supplementary Figure 14. Confusion matrices for binary classification models with threshold probability 0.5. a** Superconducting materials classification of 20% of the collected data from MPDS[11] and SuperCon[12] with respect to transition temperature 10 K; **b** magnetic materials classification of 20% of the collected data from MPDS, with respect to Curie temperature 300 K; **c** classification materials with reported values of energy band gap with respect to energy gap value 4.5 eV, test set is 20% of randomly selected data collected from MPDS.

The corresponding average accuracy, F1 score and the Matthews' correlation coefficients (MCC) are presented for the three models in Supplementary Table 3.

**Supplementary Table 3. Average binary classifications metrics of the maximum values of exhibited properties in the phase field.**

| **Metrics** | Superconducting Tc >10 K | Magnetic Tc > 300 K | Energy gap > 4.5 eV |
|---|---|---|---|
| **Accuracy, %** | 80.4 | 86.2 | 75.6 |
| **F1 score, %** | 72.9 | 84.2 | 75.3 |
| **MCC** | 0.608 | 0.711 | 0.523 |

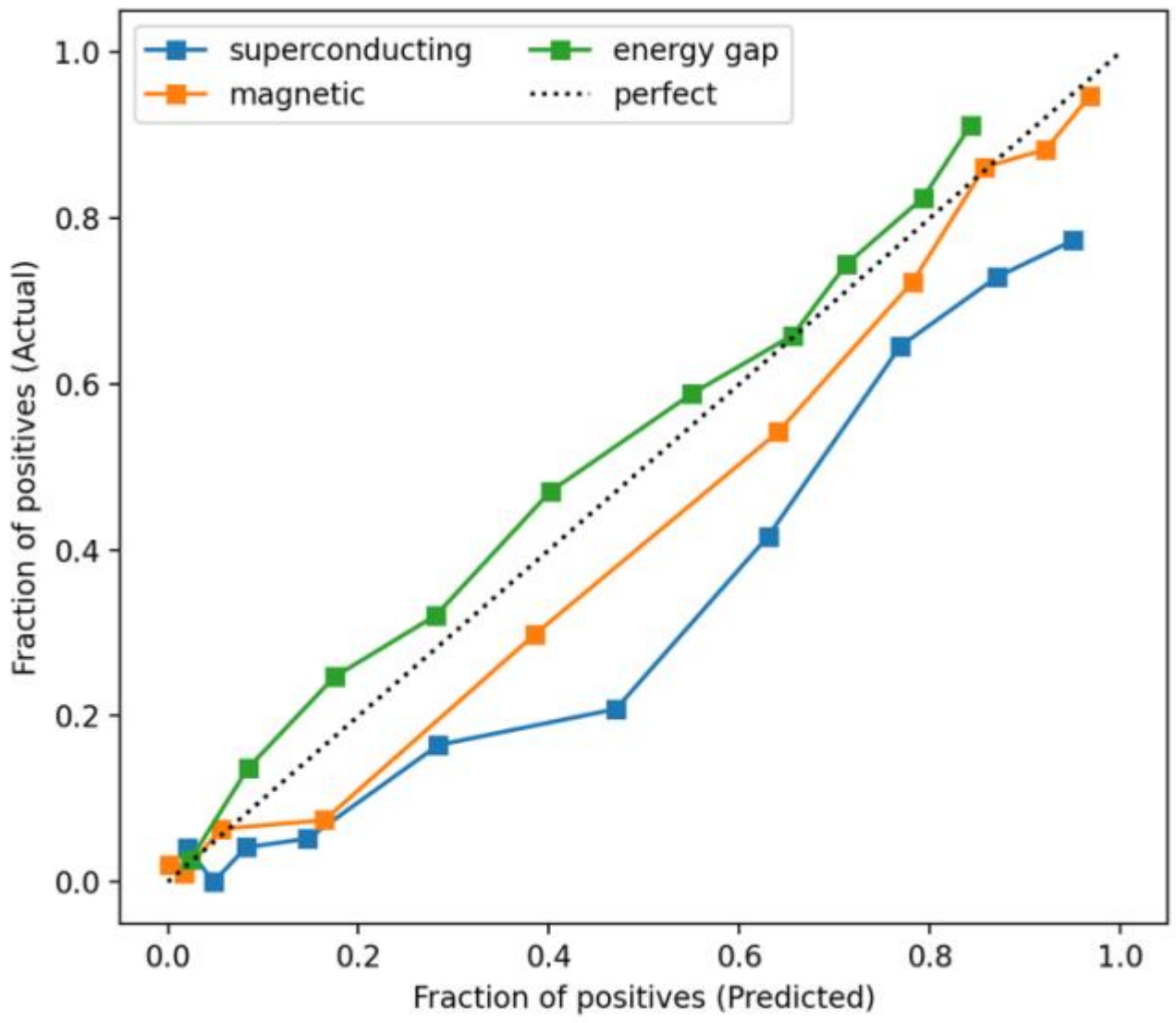

**Supplementary Figure 15. Calibration curves for the three classifiers demonstrate reliability of classification for predicting candidate phase fields with property values above the thresholds (true positives) for each dataset studied.**

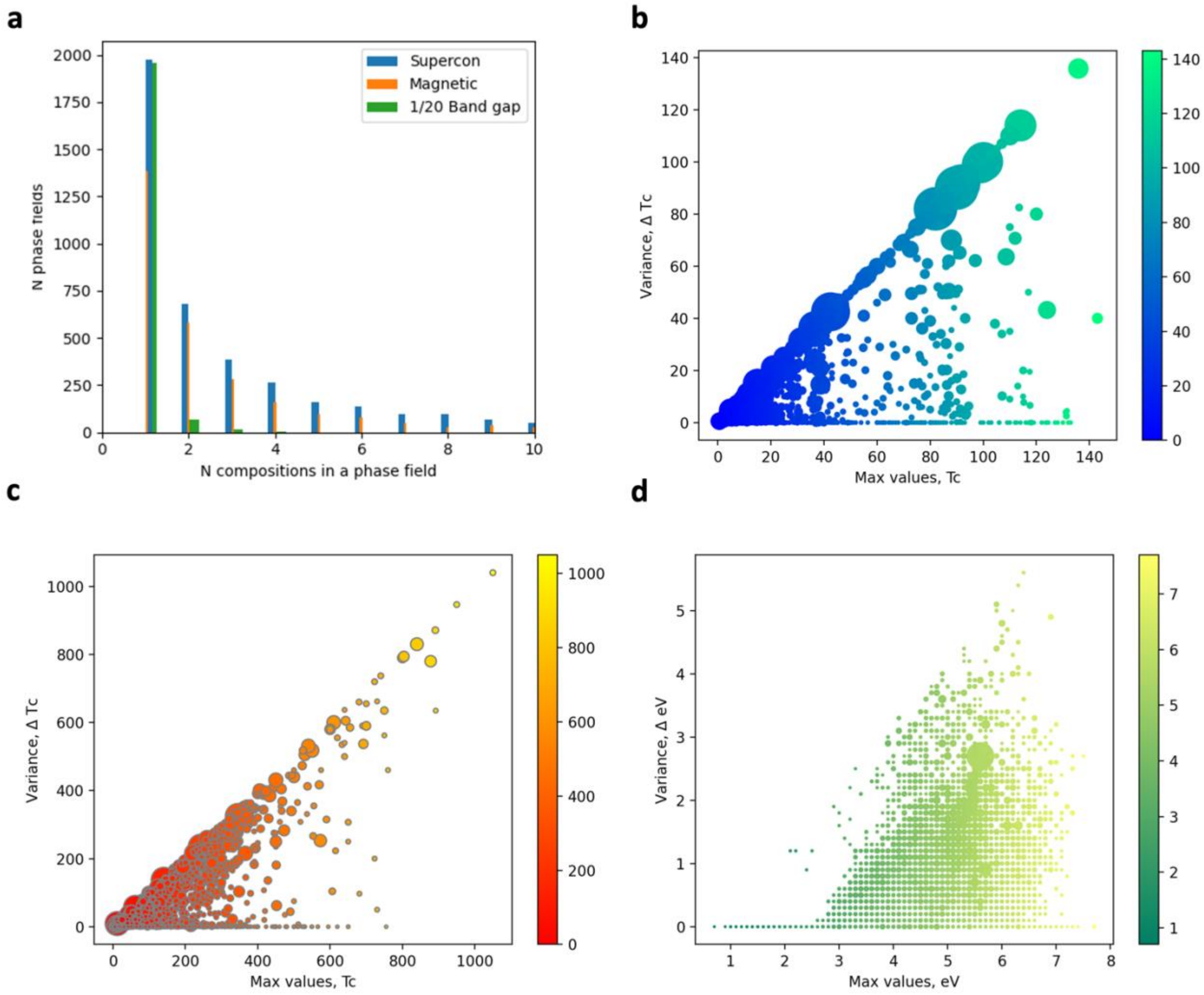

**Supplementary Figure 16. Distribution of number of compositions and properties variance in the phase fields. a** Distribution of phase fields regarding the number of representative compositions in three datasets: Majority of the phase fields are represented by single compositions only; **b** Variance in reported superconducting transition temperatures within phase fields; **c** Variance in reported Curie temperatures within phase fields; **d** Variance in reported energy band gaps within phase fields; In **b**, **c**, **d** markers sizes correspond to number of compositions reported within the phase fields; colours – to maximum values

## Combination of probabilities of high-values properties (merit probability) and synthetic uncertainties

We combine the outcomes of the classifying neural network and autoencoder to rank unexplored ternary combinations of elements. For the unexplored ternary combinations we consider all possible combinations of 87 atoms, that exclude rare and toxic elements and have sufficient data in Materials Project to be reasonably well learnt with the proposed unsupervised approach described above. The total number of ternary combinations, therefore, is $87 \times 86 \times 85 / 3! = 105995$, among them 12297 have a reported value of energy band gap in MPDS (and in peer-reviewed literature), 1953 are reported to have magnetic properties and a corresponding Curie temperature in MPDS, and 1716 are reported to have superconducting properties and a corresponding critical temperature in a combined dataset from SuperCon and MPDS.

The best ranking combinations, illustrated in Figure 4 in the main text are presented in the Supplementary Tables 4-6. Among the considered phase field there are entries that have been synthesized and reported in ICSD-v2021[13], but do not have records in MPDS and SuperCon concerning the properties studied here. These entries did not enter the training datasets and are highlighted in bold in the Supplementary Tables 4-6. These entries have been predicted to have low synthetic uncertainty, that provides experimental verification of the proposed method for ML assessment of synthetic accessibility. The full list of the predicted scores for the yet experimentally unexplored ternary phase fields can be found along with the PhaseSelect software.

**Supplementary Table 4. Predicted probabilities of the best unexplored ternary phase fields to manifest superconducting $T_c > 10$ K and their synthetic uncertainty scores. The phase fields, in which compounds are synthesized but were not included into the training data are highlighted in bold.**

| Phase fields | Probability Tc > 10 K | Synthetic uncertainty |
|---|---|---|
| **N Fe Nb** | **0.7433** | **0.1643** |
| Mg Fe As | 0.7295 | 0.1557 |
| Mg Fe Nb | 0.7278 | 0.1016 |
| **Fe As Nb** | **0.7261** | **0.0903** |
| N Cl Nb | 0.7238 | 0.1781 |
| **Fe Se Nb** | **0.7212** | **0.124** |
| N Mg Zr | 0.72 | 0.1776 |
| N K Nb | 0.7194 | 0.1798 |
| Mg V Fe | 0.7189 | 0.1589 |
| **N Na Nb** | **0.7187** | **0.1774** |
| **Ca Fe As** | **0.7162** | **0.1477** |
| **V Fe As** | **0.7154** | **0.1299** |
| **Fe Ga As** | **0.7126** | **0.1709** |
| **N Ca Nb** | **0.7111** | **0.1665** |

**Supplementary Table 5. Predicted probabilities of the best unexplored ternary phase fields to manifest Curie $T_c$ > 300 K and their synthetic uncertainty scores. The phase fields, in which compounds are synthesized[13] but were not included into the training data are highlighted in bold.**

| Phase fields | Probability $T_c$ > 300 K | Synthetic uncertainty |
|---|---|---|
| **Ti Fe Ta** | **0.7181** | **0.0658** |
| **Fe Mo Hf** | **0.717** | **0.0685** |
| **Ti Fe Hf** | **0.716** | **0.0603** |
| Fe Y Nb | 0.7159 | 0.0681 |
| Fe Y Hf | 0.7154 | 0.0557 |
| V Fe Ta | 0.7136 | 0.0631 |
| Cr Fe Ta | 0.7131 | 0.057 |
| Ti Fe Hg | 0.7123 | 0.0642 |
| **Cr Fe Zr** | **0.7123** | **0.0619** |
| **Fe Hf Ta** | **0.7117** | **0.0467** |
| **Fe Zr Hf** | **0.7113** | **0.0426** |
| Fe Nb Ta | 0.7111 | 0.0522 |
| Fe Y Hg | 0.7106 | 0.0598 |
| V Fe Nb | 0.7106 | 0.0699 |
| **V Fe Hf** | **0.7101** | **0.0575** |

**Supplementary Table 6. Predicted probabilities of the best unexplored ternary phase fields to manifest energy band gap > 4.5 eV and their synthetic uncertainty scores. The phase fields, in which compounds are synthesized but were not included into the training data[11] are highlighted in bold.**

| Phase fields | Probability $E_g$ > 4.5 eV | Synthetic uncertainty |
|---|---|---|
| **Cs F Pb** | **0.7613** | **0.8852** |
| F Hg Bi | 0.7603 | 0.0897 |
| **F Hg Pb** | **0.759** | **0.0811** |
| F Te Hf | 0.7575 | 0.0975 |
| F Y Bi | 0.7575 | 0.0984 |
| F Hf Bi | 0.7574 | 0.0906 |
| F As Hf | 0.7572 | 0.0984 |
| Cl I Hf | 0.7561 | 0.0999 |
| **F Cd Bi** | **0.7561** | **0.0882** |
| F Au Pb | 0.7556 | 0.0932 |
| F Hf Pb | 0.7556 | 0.082 |
| F Cd Pb | 0.755 | 0.0796 |
| F V Bi | 0.7531 | 0.0904 |

## Prediction of superconducting behaviour for reported phase fields in ICSD-v2021

We apply PhaseSelect ensembles of classification models to identify likely candidates for novel superconducting materials among the phase fields that have been reported to form stable compounds in ICSD-v2021, but were not investigated from the perspectives of superconducting applications and reported in MPDS and SuperCon (hence were not included into the training dataset). The excerpt of these predictions is presented in Supplementary Table 7; classification of all binary, ternary and quaternary phase field in ICSD with respect to the maximum accessible value of superconducting critical temperature is uploaded in[14].

**Supplementary Table 7. Predicted probabilities of superconducting behaviour at $T_c$ > 10 K for the best ternary phase fields reported to form stable structures in ICSD. (Excerpt for p > 0.7. The full list is in[14]).**

| Phase fields | Probability $T_c$ > 10 K | Phase fields | Probability $T_c$ > 10 K |
|---|---|---|---|
| Fe N Nb | 0.7466 | Mo N Nb | 0.7142 |
| Fe Li N | 0.7391 | C Li N | 0.7131 |
| Fe Ga N | 0.7383 | As Fe Nb | 0.7113 |
| C Fe N | 0.7352 | Al N Nb | 0.7111 |
| Fe Mo N | 0.7343 | C Ga N | 0.7102 |
| Ba Fe N | 0.733 | Ga N V | 0.7101 |
| Fe N Se | 0.7324 | N Nb V | 0.7096 |
| Ca Fe N | 0.7322 | Ca Fe O | 0.709 |
| C Mg N | 0.7305 | C N V | 0.709 |
| Fe Mg O | 0.7302 | Ba Fe O | 0.7087 |
| Li Mg N | 0.7291 | B Mg N | 0.7077 |
| Ga N Nb | 0.7262 | Fe Nb Se | 0.7075 |
| Ga Mg N | 0.7261 | C K N | 0.7075 |
| C N Nb | 0.726 | Ca Mg N | 0.7065 |
| Fe N Zr | 0.725 | As Ca Fe | 0.7058 |
| Mg Mo N | 0.7229 | C Cl N | 0.7056 |
| Fe Ga Nb | 0.7227 | N Na Nb | 0.7056 |
| Fe N Sr | 0.7226 | As Fe V | 0.7055 |
| Cl Mg N | 0.7218 | N Nb Zr | 0.7054 |
| Cu Fe O | 0.7197 | Ba N Nb | 0.7037 |
| Fe N Sn | 0.7194 | As Fe Ga | 0.7034 |
| Fe Mn N | 0.7192 | Fe N Pt | 0.7023 |
| Fe N O | 0.7186 | C N Na | 0.702 |
| As Fe O | 0.7175 | Ca N Nb | 0.7019 |
| Fe N Y | 0.7173 | As Ba Fe | 0.7017 |
| C Mo N | 0.717 | C Ca N | 0.7014 |
| Fe Ga V | 0.7157 | As Fe K | 0.7007 |
| Li N Nb | 0.7143 | | |

## Tools and Libraries

PhaseSelect[10] has been built using Python 3.7.4, Tensorflow 2.4.1, Scikit-learn 0.24.0, Numpy 0.19.2, Pandas 1.1.4. The figures in the main text and Supplementary figures are created using Matplotlib 3.3.4.

## Supplementary References